\documentclass[12pt]{article}
\usepackage[english]{babel}
\usepackage[dvips]{epsfig}
\usepackage{cite}
\usepackage{amsmath,amssymb}
\usepackage{epsfig}

\newcommand{\nn}{\nonumber}

\begin{document}
\thispagestyle{empty}
\centerline{DESY~14-102\hfill ISSN 0418-9833}
\centerline{July 2014\hfill}

\vspace*{2.0cm}

\begin{center}
 {\large \bf
HYPERDIRE \\
HYPERgeometric functions DIfferential REduction: \\
Mathematica-based packages for the differential reduction of generalized
hypergeometric functions: \\
Horn-type hypergeometric functions of two variables
 }
\end{center}
 \vspace*{0.8cm}

\begin{center}
{\sc Vladimir~V.~Bytev,$^{a,b,}$\footnote{E-mail: bvv@jinr.ru}
Bernd~A.~Kniehl$^{a,}$\footnote{E-mail: kniehl@desy.de}} \\
 \vspace*{1.0cm}
{\normalsize $^{a}$ II. Institut f\"ur Theoretische Physik, Universit\"at Hamburg,}\\
{\normalsize Luruper Chaussee 149, 22761 Hamburg, Germany} \\
\bigskip
{\normalsize $^{b}$ Joint Institute for Nuclear Research,} \\
{\normalsize $141980$ Dubna (Moscow Region), Russia}
\end{center}

\begin{abstract}
HYPERDIRE is a project devoted to the creation of a set of Mathematica-based
programs for the differential reduction of hypergeometric functions.
The current version allows for manipulations involving the full set of
Horn-type hypergeometric functions of two variables, including 30 functions. 
\end{abstract}

\newpage

{\bf\large PROGRAM SUMMARY}
\vspace{4mm}
\begin{sloppypar}
\noindent   {\em Title of program\/}: HYPERDIRE \\[2mm]
   {\em Version\/}: 1.0.0 \\[2mm]
   {\em Release\/}: 1.0.0 \\[2mm]
   {\em Catalogue number\/}: \\[2mm]
   {\em Program obtained from\/}: {\tt https://sites.google.com/site/loopcalculations/home} \\[2mm]
   {\em E-mail\/}: {\tt bvv@jinr.ru} \\[2mm]
   {\em Licensing terms\/}: GNU General Public Licence  \\[2mm]
   {\em Computers\/}: all computers running Mathematica \\[2mm]
   {\em Operating systems\/}:  operation systems running Mathematica\\[2mm]
   {\em Programming language\/}: Mathematica \\[2mm]
   {\em Nature of the problem\/}:
                  reduction of Horn-type hypergeometric functions of two variables 
                  to a set of basis functions
                \\[2mm]
   {\em Method of solution\/}: differential reduction
                \\[2mm]
   {\em Restriction on the complexity of the problem}: none \\[2mm]
   {\em Typical running time}: depending on the complexity of the problem
\end{sloppypar}
%
\newpage

\section{Introduction}
This paper describes a major extension of HYPERDIRE
\cite{hyperdire1,hyperdire2}, which is a set of Mathematica-based \cite{math}
program packages for manipulations involving Horn-type hypergeometric functions
\cite{appell,erdelyi,bateman,Gelfand} on the basis of differential equations
\cite{SST}.  
The creation of these program packages is motivated by the importance of
Horn-type hypergeometric functions for analytical evaluations of Feynman
diagrams, especially at the one-loop level \cite{one-loop}.
Possible applications of the differential-reduction algorithm to Feynman
diagrams beyond the one-loop level were discussed in Ref.~\cite{our}.

The aim of this paper is to present an extended version of HYPERDIRE, which
includes the full set of Horn-type hypergeometric functions of two variables.
Specifically, these are the 30 functions listed in Table~\ref{FunctList}.
For completeness, we present also the full list of inverse operators for the
differential reduction, in Appendix~\ref{AppendixOPERATORS}.

\section{Differential reduction}

Let us consider the hypergeometric function $H(\vec{J};\vec{z})$ depending on a
set of {\it contiguous} variables, $z_1, \cdots, z_k$, and a set of
{\it discrete} variables, $J_1, \cdots, J_r$,
and satisfying the following two additional conditions:\\
\begin{itemize}
\item
There are linear differential operators which shift the values of the 
parameters $J_a$ by $\pm 1$ (step-up and step-down operators):
\begin{equation}
R_{a,\vec{K}} \frac{\partial^{\vec{K}}}{\partial \vec{z}}
H(J_1, \cdots, J_{a-1}, J_a, J_{a+1}, \cdots, J_r;\vec{z})
=
H(J_1, \cdots, J_{a-1}, J_a \pm 1, J_{a+1}, \cdots, J_r;\vec{z}) \;,
\label{shift}
\end{equation}
where
$ 
\frac{\partial^{\vec{L}}}{\partial \vec{z}} =  
\frac{\partial^{l_1+\cdots+ l_k}}{\partial z_1^{l_1} \cdots \partial z_k^{l_k}}
$
and 
$R_{a,\vec{K}}(\vec{z})$ are rational functions.

\item
The function $H(\vec{J};\vec{z})$ satisfies the following homogeneous linear
system of partial differential equations (PDEs):
\begin{equation}
P_{\vec{L}} \frac{\partial^{\vec{L}}}{\partial \vec{z}} H(\vec{J};\vec{z}) = 0 \;,
\label{homogeneous}
\end{equation}
where 
$P_{\vec{L}}(\vec{z})$ are polynomial functions.
\end{itemize}

We assume that Eq.~(\ref{homogeneous}) may be converted to the Pfaff form,
\begin{eqnarray}
\sum_{\vec{J},k} P_{\vec{J};k}(\vec{a};\vec{z}) \frac{\partial}{\partial z_k} F(\vec{a};\vec{z}) = 0 
& \Rightarrow & 
\Biggl\{
d_k \omega_i(\vec{z}) = \Omega_{ij}^k(\vec{z}) \omega_j(\vec{z}) dz_k  \;,
\quad 
d_r \left[ d_k \omega_i(\vec{z}) \right] = 0 
\Biggr\} \;.
\label{Pfaff} 
\end{eqnarray}
Then, the differential operators inverse with respect to those defined by
Eq.~(\ref{shift}) may be constructed according to Ref.~\cite{SST}.

For certain values of the parameters, the coefficients entering the
differential operators may be equal to zero or infinity.
In this case, the result of the differential reduction may be expressed in
terms of a simpler hypergeometric function.
In Table~\ref{FunctList3}, the sets of exceptional parameters are listed for
all the Horn-type hypergeometric functions considered here.
 
Applying the direct and inverse differential operators to the hypergeometric
function $H(\vec{J};\vec{z})$, the values of the parameters $\vec{J}$ can be
shifted by arbitrary integers as
\begin{equation}
Q(\vec{z}) H(\vec{J}+\vec{m};\vec{z}) =  
\sum_{\vec{k}=0}^r Q_{\vec{k}}(\vec{z}) \frac{\partial^{\vec{k}}}{\partial \vec{z}} H(\vec{J};\vec{z}) \;,
\label{reduction}
\end{equation}
where $\vec{m}$ is a set of integers, $Q(\vec{z})$ and $Q_{\vec{J}}(\vec{z})$ are
polynomials, and $r$ is the holonomic rank of the homogeneous linear system of 
PDEs in Eq.~(\ref{homogeneous}).

Let us recall that, for a Horn-type hypergeometric function, the homogeneous
linear system of PDEs can be derived from the coefficients of the series
expansion about $\vec{z}=\vec{0}$,\footnote{%
Under special conditions depending on the values of the parameters, also the
Mellin-Barnes integral may be used for obtaining the homogeneous linear system
of PDEs \cite{beukers,KK2012}.}
$$
H = \sum_{\vec{m}} C(\vec{m}) \vec{z}^{\vec{m}}.
$$
In this case, the ratio of two coefficients may be represented as a ratio of
two polynomials,
\begin{equation}
\frac{C(\vec{m}+\vec{e}_j)}{C(\vec{m})}  =  \frac{P_j(\vec{m})}{Q_j(\vec{m})} 
\;,
\label{pre-diff}
\end{equation}
where $\vec{e}_j = (0,\cdots,0,1,0,\cdots,0)$ denotes the unit vector with
unity in its $j^{\rm th}$ component, so that the Horn-type hypergeometric
function satisfies the following homogeneous linear system of PDEs:
\begin{equation}
\left[
Q_j\left(
\sum_{k=1}^r z_k\frac{\partial}{\partial z_k}
\right)
\frac{1}{z_j}
-
 P_j\left(
\sum_{k=1}^r z_k\frac{\partial}{\partial z_k}
\right)
\right]
H(\vec{J}; \vec{z}) = 0 \;,
\label{diff}
\end{equation}
where $j=1, \ldots, r$.
%
%
%
%
\section{Horn-type hypergeometric functions of two variables}

The Horn-type hypergeometric function $H(\vec{J};z_1,z_2)$ of two variables
has the following series representation about $z_1=z_2=0$: 
\begin{eqnarray}
H(\vec{\gamma};\vec{\sigma};z_1,z_2)
=
\sum_{m_1,m_2=0}^\infty
\Biggl(
\frac{
\Pi_{a=0}^K
\Gamma\left(\mu_1^{(a)} m_1 + \mu_2^{(a)} m_2 + \gamma_a \right)
\Gamma^{-1}(\gamma_a)
}
{
\Pi_{b=0}^L
\Gamma\left(\nu_1^{(b)} m_1 + \nu_2^{(b)} m_2 + \sigma_b \right)
\Gamma^{-1}(\sigma_b)
}
\Biggr)
z_1^{m_1} z_2^{m_2} \;,
\label{H}
\end{eqnarray}
where
$\mu_j^{(a)}, \nu_k^{(b)} \in \mathbb{Z}$ and
$\gamma_j,\sigma_k \in \mathbb{C}$.
The sequences $\vec{\gamma}=(\gamma_1,\cdots, \gamma_K)$ and
$\vec{\sigma}=(\sigma_1,\cdots, \sigma_L)$ are called {\it upper} and
{\it lower} parameters, respectively.
The function $H(\vec{J};z_1,z_2)$ satisfies the following homogeneous linear
system of PDEs of second order:
\begin{eqnarray}
\theta_{11} H(\vec{J};\vec{z})
& = &
\Biggl\{
P_0(\vec{z}) \theta_{12}
+
P_1 (\vec{z}) \theta_1
+
P_2 (\vec{z}) \theta_2
+
P_3 (\vec{z})
\Biggr\}
H(\vec{J};\vec{z}) \;,
\nonumber
\\
\theta_{22} H(\vec{J};\vec{z})
& = &
\Biggl\{
R_0(\vec{z}) \theta_{12}
+
R_1 (\vec{z}) \theta_1
+
R_2 (\vec{z}) \theta_2
+
R_3 (\vec{z})
\Biggr\}
H(\vec{J};\vec{z}) \;,
\label{eq2}
\end{eqnarray}
where
$\vec{z} = (z_1,z_2)$ accommodates the two variables,
$\{P_j,R_j\}$ are rational functions,
$\theta_j = z_j \partial_{z_j}$ with $j=1,2$,
and
$
\theta_{i_1\cdots i_k} =
\theta_{i_i} \cdots \theta_{i_k}.
$
It is well known \cite{appell} that, under the condition
$
1 - P_0(\vec{z}) R_0(\vec{z}) = 0,
$
Eq.~(\ref{eq2}) can be reduced to the Pfaff system in Eq.~(\ref{Pfaff})
of three PDEs
\begin{equation}
d \vec{f} = R \vec{f} \;,
\label{pfaff4}
\end{equation}
where
$
\vec{f} = \left( H(\vec{J};\vec{z}), \theta_1 H(\vec{J};\vec{z}), \theta_2 H(\vec{J};\vec{z}) \right)\;.
$
In this case, Eq.~(\ref{eq2}) has
three  solutions, 
and Eq.~(\ref{reduction}) takes the following form: 
\begin{equation}
Q(\vec{z}) H(\vec{J}+\vec{m};\vec{z}) =  
Q_0(\vec{z})  H(\vec{J};\vec{z})
+ 
Q_1(\vec{z})  \theta_1 H(\vec{J};\vec{z})
+ 
Q_2(\vec{z})  \theta_2 H(\vec{J};\vec{z})
\;,
\label{eq3funct}
\end{equation}
where $\vec{m}$ is a set of integers and $Q(\vec{z})$, $Q_0(\vec{z})$,
$Q_1(\vec{z})$, and $Q_2(\vec{z})$ are polynomials.

In the case
$
1 - P_0(\vec{z}) R_0(\vec{z}) \neq 0,
$
Eq.~(\ref{eq2}) has four independent
solutions and may be reduced to the Pfaff system of four PDEs in
Eq.~(\ref{pfaff4}) with 
$
\vec{f} = \left( H(\vec{J};\vec{z}), \theta_1 H(\vec{J};\vec{z}), \theta_2 H(\vec{J};\vec{z}), 
                 \theta_{12} H(\vec{J};\vec{z}) \right).
$
In this case, Eq.~(\ref{reduction}) has the following form: 
\begin{equation}
Q(\vec{z}) H(\vec{J}+\vec{m};\vec{z}) =  
Q_0(\vec{z})  H(\vec{J};\vec{z})
+ 
Q_1(\vec{z})  \theta_1 H(\vec{J};\vec{z})
+ 
Q_2(\vec{z})  \theta_2 H(\vec{J};\vec{z})
+ 
Q_{12}(\vec{z})  \theta_{12} H(\vec{J};\vec{z}) \;,
\label{eq4funct}
\end{equation}
where $\vec{m}$ is a set of integers and
$Q(\vec{z})$, $Q_0(\vec{z})$, $Q_1(\vec{z})$, $Q_2(\vec{z})$, and
$Q_{12}(\vec{z})$ are polynomials. 
In both cases, Eqs.~(\ref{eq3funct}) and (\ref{eq4funct}), the construction of
the inverse differential operators reduces to the construction of some inverse
matrices, of the $3\times3$ and $4\times4$ types, respectively, with non-zero
determinants. 
However, as was shown in Ref.~\cite{erdelyi2}, for some Horn-type
hypergeometric functions, one of the four particular solutions under the
condition 
$1 - P_0(\vec{z}) R_0(\vec{z}) \neq 0$ is a Puiseux monomial in the neighborhood
of the point $z_1=z_2=0$.
Examples include the functions $G_3$, $H_3$, $H_6$, $H_6$ (confluent), and
$H_8$ (confluent).
More details may be found in Ref.~\cite{erdelyi}.
In applications to Feynman diagrams, such solutions correspond to diagrams
which are exactly expressible in terms of Gamma functions and are typically
associated with tadpoles \cite{KK2012}.
In this case, the determinant of the corresponding matrix is equal to zero, 
and the differential reduction has the form of Eq.~(\ref{eq3funct}).
To complete the differential reduction in this case, it is necessary to
generate one PDE in addition to Eq.~(\ref{eq2}).
A detailed analysis of such systems of PDEs for Horn-type
hypergeometric functions of two variables was performed in Ref.~\cite{miller}.
The most systematic analysis of the criteria for the existence of such types
of solutions for A-hypergeometric systems \cite{Gelfand} was presented in
Ref.~\cite{dickenstein}.

In Table~\ref{FunctList2}, the locus of singularities of the homogeneous linear
system of PDEs of second order with two variables defined by Eq.~(\ref{eq2}) is
specified for each of the Horn-type hypergeometric functions considered here.

\section{HornFunctions --- Mathematica-based program for the differential
reduction of 30 Horn-type hypergeometric functions}

In this section, we present the Mathematica-based\footnote{%
It was tested using Mathematica~8.0.}
program package {\bf HornFunctions} for the differential reduction of the 30
Horn-type hypergeometric functions of two variables. 
They are listed in Table~\ref{FunctList}.
The differential reduction of the Appell functions, namely $F_1$, $F_2$, $F_3$,
and $F_4$, was implemented in the program package {\bf AppellF1F4}
\cite{hyperdire2}.  

For the Horn-type hypergeometric functions defined in Eq.~(\ref{H}), the direct
differential operators for the upper and lower parameters were constructed
in Ref.~\cite{contiguous}.
For the upper parameters, they have the following form:  
\begin{eqnarray}
H(\vec{\gamma}+\vec{e_a};\vec{\sigma};\vec{z})
= 
\frac{1}{\gamma_a}
\left(\mu_1^{(a)} \theta_1 \!+\! \mu_2^{(a)} \theta_2 \!+\! \gamma_a \right)
H(\vec{\gamma};\vec{\sigma};\vec{z}) \;.
\label{do1}
\end{eqnarray}
Similar relations also exist for the lower parameters:
\begin{eqnarray}
H(\vec{\gamma};\vec{\sigma}-\vec{e}_b;\vec{z})
& = &
\frac{1}{\sigma_b \!-\! 1}
\left(\nu_1^{(b)} \theta_1 \!+\! \nu_2^{(b)} \theta_2 \!+\! \sigma_b-1 \right)
H(\vec{\gamma};\vec{\sigma};\vec{z})
\;.
\label{do2}
\end{eqnarray}
The program package {\bf HornFunctions} allows one to automatically perform the
differential reduction in accordance with Eq.~(\ref{reduction}).
It is freely available from Ref.~\cite{bytev:hyper}.
Its current version, 1.0, only handles non-exceptional values of the
parameters.

\begin{table}[H]
\caption{\label{FunctList}%
List of Horn-type hypergeometric functions implemented in the program package
{\bf HornFunctions} and their respective operators.}
\begin{center}
\begin{tabular}{|c|c|}
\hline
Horn-type function & HYPERDIRE \\
\hline
$G_1(a,b_1,b_2,z_1,z_2)$ &  G1IndexChange[...] \\
$G_2(a_1,a_2,b_1,b_2,z_1,z_2)$ &  G2IndexChange[...] \\
$G_3(a_1,a_2,z_1,z_2)$ &  G3IndexChange[...] \\
$H_1(a,b,c,d,z_1,z_2)$ &  H1IndexChange[...] \\
$H_2(a,b,c,d,e,z_1,z_2)$ &  H2IndexChange[...] \\
$H_3(a,b,c,z_1,z_2)$ &  H3IndexChange[...] \\
$H_4(a,b,c,d,z_1,z_2)$ &  H4IndexChange[...] \\
$H_5(a,b,c,z_1,z_2)$ &  H5IndexChange[...] \\
$H_6(a,b,c,z_1,z_2)$ &  H6IndexChange[...] \\
$H_7(a,b,c,d,z_1,z_2)$ &  H7IndexChange[...] \\
\hline
\multicolumn{2}{|c|}{confluent series:} \\
\hline
$\Phi_1(a,b,c,z_1,z_2)$ &  Phi1IndexChange[...] \\
$\Phi_2(b_1,b_2,z_1,z_2)$ &  Phi2IndexChange[...] \\
$\Phi_3(b,c,z_1,z_2)$ &  Phi3IndexChange[...] \\

$\Psi_1(a,b,c_1,c_2,z_1,z_2)$ &  Psi1IndexChange[...] \\
$\Psi_2(a,b,c,z_1,z_2)$ &  Psi2IndexChange[...] \\

$\Theta_1(a_1,a_2,b,c,z_1,z_2)$ &  Theta1IndexChange[...] \\
$\Theta_2(a,b,c,z_1,z_2)$ &  Theta2IndexChange[...] \\

$\Gamma_1(a,b_1,b_2,c,z_1,z_2)$ &  Gamma1IndexChange[...] \\
$\Gamma_2(b_1,b_2,c,z_1,z_2)$ &  Gamma2IndexChange[...] \\

$H_1(a,b,c,z_1,z_2)$ &  H1cIndexChange[...] \\
$H_2(a,b,c,d,z_1,z_2)$ &  H2cIndexChange[...] \\
$H_3(a,b,c,z_1,z_2)$ &  H3cIndexChange[...] \\
$H_4(a,c,d,z_1,z_2)$ &  H4cIndexChange[...] \\
$H_5(a,b,z_1,z_2)$ &  H5cIndexChange[...] \\
$H_6(a,c,z_1,z_2)$ &  H6cIndexChange[...] \\
$H_7(a,c,d,z_1,z_2)$ &  H7cIndexChange[...] \\
$H_8(a,b,z_1,z_2)$ &  H8cIndexChange[...] \\
$H_9(a,b,c,z_1,z_2)$ &  H9cIndexChange[...] \\
$H_{10}(a,b,z_1,z_2)$ &  H10cIndexChange[...] \\
$H_{11}(a,b,c,d,z_1,z_2)$ &  H11cIndexChange[...] \\
\hline
\end{tabular}
\end{center}
\end{table} 


\begin{table}[H]
\caption{\label{FunctList3}%
Sets of exceptional parameters for the Horn-type hypergeometric functions
implemented in the program package {\bf HornFunctions}.}
\begin{center}
\begin{tabular}{|c|c|}
\hline
Horn-type function & set of exceptional parameters  \\
\hline
$G_1(a,b_1,b_2,z_1,z_2)$ & $\{ a,a+b_i,b_1+b_2 \} \in \mathbb{Z} $ \\
$G_2(a_1,a_2,b_1,b_2,z_1,z_2)$ &  $\{a_i, a_i+b_i,b_1+b_2 \} \in \mathbb{Z} $ \\
$G_3(a_1,a_2,z_1,z_2)$ &   $\{ 2a_1+a_2,2a_2+a_1 \} \in \mathbb{Z} $ \\
$H_1(a,b,c,d,z_1,z_2)$ &  $\{b,c,a+b,a+c,a+b-2d,a+c-d,b-d \} \in \mathbb{Z} $  \\
$H_2(a,b,c,d,e,z_1,z_2)$ &  $\{b,c,d,a+c,a+d,a+c-e,a+d-e,b-e \} \in \mathbb{Z} $  \\
$H_3(a,b,c,z_1,z_2)$ &  $\{a,b,a-c,a+b-2c \} \in \mathbb{Z} $ \\
$H_4(a,b,c,d,z_1,z_2)$ &  $\{a,b,a-2c,a-d,a-2c-d,b-d \} \in \mathbb{Z} $\\
$H_5(a,b,c,z_1,z_2)$ &  $\{a,a+2b,a-c,a+2b-3c  \} \in \mathbb{Z} $ \\
$H_6(a,b,c,z_1,z_2)$ &  $\{c,a+b,a+c,a+2b  \} \in \mathbb{Z} $ \\
$H_7(a,b,c,d,z_1,z_2)$ & $\{b,c,a+b,a+c,a+b-2d,a+c-2d  \} \in \mathbb{Z} $ \\
\hline
\multicolumn{2}{|c|}{confluent series:} \\
\hline
$\Phi_1(a,b,c,z_1,z_2)$ &  $\{a,b,a-c  \} \in \mathbb{Z} $ \\
$\Phi_2(b_1,b_2,z_1,z_2)$ &  $\{b_i,b_1+b_2-c \} \in \mathbb{Z} $ \\
$\Phi_3(b,c,z_1,z_2)$ &  $\{b \} \in \mathbb{Z} $ \\

$\Psi_1(a,b,c_1,c_2,z_1,z_2)$ & $\{a,b,a-c_i,a-c_1-c_2,b-c_1 \} \in \mathbb{Z} $ \\
$\Psi_2(a,b,c,z_1,z_2)$ &   $\{a,a-c_i,a-c_1-c_2 \} \in \mathbb{Z} $\\

$\Theta_1(a_1,a_2,b,c,z_1,z_2)$ &  $\{a_i,b,a_1+a_2-c,b+a_2-c \} \in \mathbb{Z} $ \\
$\Theta_2(a,b,c,z_1,z_2)$ &  $\{a,b \} \in \mathbb{Z} $ \\

$\Gamma_1(a,b_1,b_2,c,z_1,z_2)$ &  $\{a,a+b_1,b_1+b_2 \} \in \mathbb{Z} $ \\
$\Gamma_2(b_1,b_2,c,z_1,z_2)$ &   $\{b_1+b_2 \} \in \mathbb{Z} $ \\

$H_1(a,b,c,z_1,z_2)$ &   $\{b,a+b,a+b-2c,b-c \} \in \mathbb{Z} $ \\
$H_2(a,b,c,d,z_1,z_2)$ & $\{b,c,a+c,a+c-d,b-d \} \in \mathbb{Z} $ \\
$H_3(a,b,c,z_1,z_2)$ &  $\{b,b-c \} \in \mathbb{Z} $ \\
$H_4(a,c,d,z_1,z_2)$ &   $\{b,a+b,a+b-c \} \in \mathbb{Z} $\\
$H_5(a,b,z_1,z_2)$ &   \\
$H_6(a,c,z_1,z_2)$ & $\{a,a-b \} \in \mathbb{Z} $ \\
$H_7(a,c,d,z_1,z_2)$ &   $\{a,a-d,a-2c,a-2c-d \} \in \mathbb{Z} $\\
$H_8(a,b,z_1,z_2)$ &  $\{a+b,a+2b \} \in \mathbb{Z} $ \\
$H_9(a,b,c,z_1,z_2)$ &  $\{b,a+b,a+b-2c \} \in \mathbb{Z} $ \\
$H_{10}(a,b,z_1,z_2)$ &   \\
$H_{11}(a,b,c,d,z_1,z_2)$ &  $\{b,c,a+b,a+c,a+b-d,a+c-d \} \in \mathbb{Z} $ \\
\hline
\end{tabular}
\end{center}
\end{table} 


\begin{table}[H]
\caption{\label{FunctList2}%
Loci of singularities of the homogeneous linear systems of PDEs of second
order with two variables for the Horn-type hypergeometric functions implemented
in the program package {\bf HornFunctions}.}
\begin{center}
\begin{tabular}{|c|c|}
\hline
Horn-type function & singularity surfaces  \\
\hline
$G_1(a,b_1,b_2,z_1,z_2)$ & $ \cup_{i=1}^2 \{z_i=0\} \cup \{ 1+z_1+z_2=0 \} \cup \{ 4z_1z_2=1\}$  \\
$G_2(a_1,a_2,b_1,b_2,z_1,z_2)$ & $ \cup_{i=1}^2 \{z_i=0\} \cup_{i=1}^2 \{ 1+z_i=0 \} \cup \{ z_1z_2=1\}$ \\
$G_3(a_1,a_2,z_1,z_2)$ &  $ \cup_{i=1}^2 \{z_i=0\} \cup \{ -1-4z_1-4z_2-18 z_1z_2+27z_1^2z_2^2=0 \}$\\
$H_1(a,b,c,d,z_1,z_2)$ &  $ \cup_{i=1}^2 \{z_i=0\} \cup_{i=1}^2 \{ 1-z_i=0 \} \cup \{ 1+z_2=0 \} $ \\
$H_2(a,b,c,d,e,z_1,z_2)$ & $ \cup_{i=1}^2 \{z_i=0\} \cup \{ 1-z_1=0 \} \cup \{ 1+z_2=0 \} $ \\
$H_3(a,b,c,z_1,z_2)$ & $ \cup_{i=1}^2 \{z_i=0\} \cup \{ 1-4z_1=0 \} \cup \{ z_1-z_2+z_2^2=0 \} $ \\
$H_4(a,b,c,d,z_1,z_2)$ &  $ \cup_{i=1}^2 \{z_i=0\} \cup \{ 1-4z_1=0 \} \cup \{ 1-z_2=0 \} $  \\
$H_5(a,b,c,z_1,z_2)$ &  $ \cup_{i=1}^2 \{z_i=0\} \cup \{ -1+z_2+z_1(-4+6z_2)=0 \}$  \\
$H_6(a,b,c,z_1,z_2)$ & $ \cup_{i=1}^2 \{z_i=0\} \cup \{ 1+4z_1=0 \} \cup\{ 1+z_2-z_1z_2^2=0\}$  \\
$H_7(a,b,c,d,z_1,z_2)$ & $ \cup_{i=1}^2 \{z_i=0\} \cup \{ -1+4z_1=0 \} \cup\{ 1+z_2=0\}$ \\
\hline
\multicolumn{2}{|c|}{confluent series:} \\
\hline
$\Phi_1(a,b,c,z_1,z_2)$ & $ \cup_{i=1}^2 \{z_i=0\} \cup \{ -1+z_1=0 \} $ \\
$\Phi_2(b_1,b_2,z_1,z_2)$ &  $\cup_{i=1}^2 \{z_i=0\} \cup \{ z_1-z_2=0 \} $\\
$\Phi_3(b,c,z_1,z_2)$ &  $ \cup_{i=1}^2 \{z_i=0\} $ \\

$\Psi_1(a,b,c_1,c_2,z_1,z_2)$ &  $ \cup_{i=1}^2 \{z_i=0\} \cup \{ -1+z_1=0 \} $ \\
$\Psi_2(a,b,c,z_1,z_2)$ &  $ \cup_{i=1}^2 \{z_i=0\} $ \\

$\Theta_1(a_1,a_2,b,c,z_1,z_2)$ &  $ \cup_{i=1}^2 \{z_i=0\} \cup \{ -1+z_1=0 \} $  \\
$\Theta_2(a,b,c,z_1,z_2)$ &  $ \cup_{i=1}^2 \{z_i=0\} \cup \{ -1+z_1=0 \} $  \\

$\Gamma_1(a,b_1,b_2,c,z_1,z_2)$ &  $ \cup_{i=1}^2 \{z_i=0\} \cup \{ 1+z_1=0 \} $ \\
$\Gamma_2(b_1,b_2,c,z_1,z_2)$ &  $ \cup_{i=1}^2 \{z_i=0\} $ \\

$H_1(a,b,c,z_1,z_2)$ & $ \cup_{i=1}^2 \{z_i=0\} \cup \{ -1+z_1=0 \} $ \\
$H_2(a,b,c,d,z_1,z_2)$ &  $ \cup_{i=1}^2 \{z_i=0\} \cup \{ -1+z_1=0 \} $  \\
$H_3(a,b,c,z_1,z_2)$ &  $ \cup_{i=1}^2 \{z_i=0\} \cup \{ -1+z_1=0 \} $ \\
$H_4(a,c,d,z_1,z_2)$ &  $ \cup_{i=1}^2 \{z_i=0\} $ \\
$H_5(a,b,z_1,z_2)$ &  $ \cup_{i=1}^2 \{z_i=0\} $ \\
$H_6(a,c,z_1,z_2)$ &  $ \cup_{i=1}^2 \{z_i=0\} \cup \{ -1+4z_1=0 \} $\\
$H_7(a,c,d,z_1,z_2)$ &  $ \cup_{i=1}^2 \{z_i=0\} \cup \{ -1+4z_1=0 \} $ \\
$H_8(a,b,z_1,z_2)$ &  $ \cup_{i=1}^2 \{z_i=0\} \cup \{ 1+4z_1=0 \} $ \\
$H_9(a,b,c,z_1,z_2)$ &  $ \cup_{i=1}^2 \{z_i=0\} \cup \{ -1+4z_1=0 \} $\\
$H_{10}(a,b,z_1,z_2)$ & $ \cup_{i=1}^2 \{z_i=0\} \cup \{ -1+4z_1=0 \} $\\
$H_{11}(a,b,c,d,z_1,z_2)$ &  $ \cup_{i=1}^2 \{z_i=0\} \cup \{ 1+z_2=0 \} $ \\
\hline
\end{tabular}
\end{center}
\end{table}

\subsection{Input format}

The program may be loaded in the standard way:
$$
<< \mathrm{"HornFunctions.m"}
$$
It includes the following basic routine for each Horn-type hypergeometric
function:
\begin{eqnarray}
{\bf \{HornName\}IndexChange}[\mbox{changingVector}, \mbox{parameterVector}],
\label{input:f1}
\end{eqnarray}
where ${\bf \{HornName\}IndexChange}$ defines the name of the Horn-type
hypergeometric function to be modified, e.g.\ ${\bf H1IndexChange}$ for the
function $H_1(\alpha,\beta,\gamma,\delta,z_1,z_2)$,\break
``${\rm parameterVector}$'' defines the list of parameters of that function,
and ``${\rm changingVector}$'' defines the set of integers by which the values
of these parameters are to be shifted, i.e.\ the vector $\vec{m}$ in
Eq.~(\ref{reduction}).
For example, the operator
\begin{eqnarray}
{\bf H1IndexChange}[\{1,-1,0,0\},\{\alpha,\beta,\gamma,\delta,z_1,z_2\}]
\label{input:f4}
\end{eqnarray}
shifts the arguments of the function
$H_1(\alpha,\beta,\gamma,\delta,z_1,z_2)$ so as to generate
$H_1(\alpha+1,\beta-1,\gamma,\delta,z_1,z_2)$.

\subsection{Output format}

The output structure of all the operators of the program package
{\bf HornFunctions} in Eq.~(\ref{input:f1}) is as follows:
\begin{eqnarray}
\{\{Q_0,Q_1,Q_2,Q_{12} \},\{ {\rm parameterVectorNew} \} \},
\end{eqnarray}
where
``${\rm parameterVectorNew}$'' is the new set of parameters,
$\vec{J}+\vec{m}$, of the function ${\bf HornName}$ and
$Q_0$, $Q_1$, $Q_2$, and $Q_{12}$ are the rational coefficient functions of the
differential operator in Eq.~(\ref{eq4funct}),
\begin{eqnarray}
{\bf HornFunct}(old param)=(Q_0+Q_1 \theta_1+Q_2\theta_2+Q_{12}\theta_1\theta_2)
{\bf HornFunct}(new param)
\end{eqnarray}
In the case of three PDEs, which corresponds to Eq.~(\ref{eq3funct}), the last
coefficient, $Q_{12}$, is identically zero.

\subsection{Examples}

{\bf Example 1}:\footnote{%
All functions in the program package HYPERDIRE generate output without
additional simplification for maximum efficiency of the algorithm.
To get the output in a simpler form, we recommend to use the command
{\bf Simplify} in addition.
\label{simpl}}
Reduction of the Horn-type hypergeometric function $G_1(a,b_1,b_2,z_1,z_2)$.
\\
\\
{\bf G1IndexChange[}\{$-1$,$-1$,$0$\},\,\{$a$,$b1$,$b2$,$z1$,$z2$\}{\bf ]}
\\
\\
\scriptsize
\begin{eqnarray}
\left\{\left\{\frac{\text{\it b1} (\text{\it z1}+1)+(\text{\it b2}-1) \text{\it z1}-1}{(\text{\it b1}-1) (\text{\it z1}+\text{\it z2}+1)},\frac{-a+\text{\it z1}
   (\text{\it b1}+\text{\it b2}-1)+1}{(a-1) (\text{\it b1}-1) (\text{\it z1}+\text{\it z2}+1)},\frac{a+\text{\it b1} \text{\it z1}+\text{\it b1}+\text{\it b2}
   \text{\it z1}+\text{\it b2}-\text{\it z1}-2}{(a-1) (\text{\it b1}-1)
   (\text{\it z1}+\text{\it z2}+1)},0\right\},  \right.
  \nn  \\ \left. \nn
   \{a-1,\text{\it b1}-1,\text{\it b2},\text{\it z1},\text{\it z2}\}\right\}
\end{eqnarray}
\normalsize
\\
As the function $G_1(a,b_1,b_2,z_1,z_2)$ only satisfies three independent PDEs,
it may be written without the mixed derivative
$\theta_1\theta_2$, as:
\begin{eqnarray}
G_1 &&   \!\!\!\!\!\!\!\!\!(a,b_1,b_2;z_1,z_2)
=
\nonumber \\ &&
\Biggl[
\frac{\left(b_2-1\right) z_1+b \left(z_1+1\right)-1}{(b-1) \left(z_1+z_2+1\right)}
+\frac{-a+\left(b+b_2-1\right)z_1+1}{(a-1) (b-1) \left(z_1+z_2+1\right)}\theta_1
\nonumber \\ &&
+\frac{a+b z_1+b_2 z_1+b+b_2-z_1-2}{(a-1) (b-1)\left(z_1+z_2+1\right)}\theta_2
\Biggr]
G_1(a-1,b_1-1,b_2; z_1,z_2).
\label{eq:g1}
\end{eqnarray}
Using Ref.~\cite{bytev:hyper}, Eq.~(\ref{eq:g1}) may be checked numerically in
specific examples.
\\
\\
{\bf Example 2}: Reduction of the Horn-type hypergeometric function
$H_1(a,b,c,d,z_1,z_2)$.
\\
\\
 {\bf H1IndexChange[}\{$-1$,$0$,$0$,$1$\},\,\{$a$,$b$,$c$,$d$,z$1$,z$2$\}{\bf ]}
\\
\\
\scriptsize
\begin{eqnarray}
\left\{\left\{\frac{(a-1) d (\text{\it z1}-1) (\text{\it z2}+1)-b \text{\it z1} (a \text{\it z2}+a+c \text{\it z2}-\text{\it z2}-1)}{(a-1) d (\text{\it z1}-1)
   (\text{\it z2}+1)},\frac{-a (\text{\it z2}+1)-b \text{\it z1} (\text{\it z2}+1)-c \text{\it z1} \text{\it z2}+d \text{\it z1} \text{\it z2}+d
   \text{\it z1}+\text{\it z2}+1}{(a-1) d (\text{\it z1}-1) (\text{\it z2}+1)},   \right.\right.  
   \nn\\
 \left.\left. \nn
   \frac{\text{\it z1} (b-\text{\it z2} (a+c-1))-d (\text{\it z1}-1)
   (\text{\it z2}+1)}{(a-1) d (\text{\it z1}-1) (\text{\it z2}+1)},\frac{-2 \text{\it z1} \text{\it z2}+\text{\it z2}+1}{(a-1) d (\text{\it z1}-1)
   (\text{\it z2}+1)}\right\},
   \{a-1,b,c,d+1,\text{\it z1},\text{\it z2}\}\right\}
\end{eqnarray}
\normalsize
\\
\\
This corresponds to the following mathematical formula:
\begin{eqnarray}
H_1(a,b,c,d;z_1,z_2)\!\!\!\!\!
&&=
\nonumber \\ &&
\Biggl[
\frac{(a-1) d \left(z_1-1\right) \left(z_2+1\right)-b z_1 \left(a z_2+a+c z_2-z_2-1\right)}{(a-1) d \left(z_1-1\right)
   \left(z_2+1\right)}
   \nn\\
   &&
   +\frac{-a \left(z_2+1\right)-b \left(z_2+1\right) z_1-c z_2 z_1+d z_1+d z_2 z_1+z_2+1}{(a-1) d
   \left(z_1-1\right) \left(z_2+1\right)}\theta_1
   \nn\\
   &&
   +\frac{z_1 \left(b-z_2 (a+c-1)\right)-d \left(z_1-1\right) \left(z_2+1\right)}{(a-1) d
   \left(z_1-1\right) \left(z_2+1\right)}\theta_2
   \nn\\
   &&
   +\frac{-2 z_1 z_2+z_2+1}{(a-1) d \left(z_1-1\right)
   \left(z_2+1\right)}\theta_1\theta_2
   \Biggr]
H_1(a-1,b,c+1,d; z_1,z_2).
\end{eqnarray}
\\
\\
{\bf Example 3}: Reduction of the confluent Horn-type hypergeometric function
$H_1(a,b,c,z_1,z_2)$.
\\
\\
 {\bf H1cIndexChange[}\{$0$,$1$,$1$\},\,\{$a$,$b$,$c$,z$1$,z$2$\}{\bf ]}
\\
\\
\scriptsize
\begin{eqnarray}
\left\{\left\{\frac{a^2 (-\text{\it z1})-a b \text{\it z1}+a c+b c+2 b \text{\it z1} \text{\it z2}-c \text{\it z2}+2 \text{\it z1} \text{\it z2}}{a c+b c},\frac{a
   (-\text{\it z1})+a-b \text{\it z1}+b+2 \text{\it z1} \text{\it z2}-\text{\it z2}}{a c+b c},\frac{\text{\it z1} (a+b+2 \text{\it z2})-c}{c (a+b)},-\frac{1}{a
   c+b c}\right\}, \right.
   \nn\\
   \nn \left.
   \{a,b+1,c+1,\text{\it z1},\text{\it z2}\}\right\}
\end{eqnarray}
\normalsize
\\
\\
This corresponds to the following mathematical formula:
\begin{eqnarray}
H_1(a,b,c;z_1,z_2)=
\Biggl[
&&\frac{a^2 \left(-z_1\right)-a b z_1+a c+b c+2 b z_1 z_2-c z_2+2 z_1 z_2}{a c+b c},
\nn\\
&&\frac{a \left(-z_1\right)+a-b z_1+b+2
   z_1 z_2-z_2}{a c+b c},\frac{z_1 \left(a+b+2 z_2\right)-c}{c (a+b)},
   \nn\\
 &&  -\frac{1}{a c+bc}
   \Biggr]
H_1(a,b+1,c+1; z_1,z_2).
\end{eqnarray}

\section{Conclusions}
\label{conclusion}

The differential-reduction algorithm \cite{SST} allows one to relate Horn-type
hypergeometric functions with parameters whose values differ by integers.
In this paper, we presented an extended version of the Mathematica-based
program package HYPERDIRE \cite{hyperdire1,hyperdire2}
for the differential reduction of generalized hypergeometric functions of Horn
type with two variables to a set of basis functions.

\section*{Acknowledgements}

We are grateful to M.Yu.~Kalmykov for fruitful discussions, useful remarks, and
valuable contributions to this paper. 
The work of V.V.B. was supported in part by the Russian Foundation for Basic
Research RFFI through Grant No.~12-02-31703 and by the Heisenberg-Landau
Program.
This work was supported in part by the German Federal Ministry for Education
and Research BMBF through Grant No.\ 05H12GUE and by the German Research
Foundation DFG through the Collaborative Research Centre No.~676
{\it Particles, Strings and the Early Universe---The Structure of Matter and
Space-Time}.

\appendix

\section{Inverse differential operators}
\label{AppendixOPERATORS}

In this appendix, we present the full list of differential operators inverse to
those defined by Eqs.~(\ref{do1}) and (\ref{do2}), which shift the upper and
lower parameters of the Horn-type hypergeometric functions, respectively.
The corresponding results for the Appell hypergeometric functions, $F_1$,
$F_2$, $F_3$, and $F_4$, were presented in Ref.~\cite{hyperdire2}.
The sets of upper and lower parameters are uniquely defined by the series
representation of the Horn-type hypergeometric function in Eq.~(\ref{H}).
In the remainder of this paper, we adopt the following notations.
For each parameter $a_i$ of the Horn-type hypergeometric function $F_k$, we
denote the four coefficient functions appearing in the differential operators
inverse to those defined by Eqs.~(\ref{do1}) and (\ref{do2}) by $A_{a_i,F_k}$,
$B_{a_i,F_k}$, $C_{a_i,F_k}$, and $D_{a_i,F_k}$.
Specifically, we have
\begin{equation}
F_k(\ldots,a_i,\ldots)=
(A_{a_i,F_k}+ B_{a_i,F_k}\theta_1+C_{a_i,F_k}\theta_2+D_{a_i,F_k}\theta_1\theta_2)
F_k(\ldots,a_i\pm1,\ldots)
\end{equation}
for upper and lower parameters, respectively.
When only three non-trivial solutions exist, we put explicitly $D_{a_i,F_k}=0$.
In the special cases when one of the four solutions is a Puiseux monomial, we
also present the extra PDE.

\subsection{Non-confluent Horn-type hypergeometric functions}

\boldmath
\subsubsection{Function $\left.G_1(a,b_1,b_2,z_1,z_2\right)$}
\unboldmath

\begin{eqnarray}
\left.G_1\text{(a,}b_1,b_2,z_1,z_2\right)
=\sum_{n_1,n_2}
\frac{(a)_{n_1+n_2} \left(b_1\right)_{n_2-n_1} \left(b_2\right)_{n_1-n_2}}{n_1! n_2!}
z_1^{n_1} z_2^{n_2}.
\end{eqnarray}

The additional PDE reads:
\begin{eqnarray}
&&\frac{\left(z_1+z_2+1\right) \left(4 z_1 z_2-1\right)}{z_1 z_2}\theta _1 \theta _2
G_1(\vec{\gamma};z_1,z_2)=
\nn\\
&&( \left(-a^2 z_1-a^2 z_2-a^2+a b_2 z_1+a b_1 z_2 
+2 a b_1 z_1 z_2+2 a b_2 z_1 z_2-a z_1-a z_2-a\right)
\nn\\
&&+\left(-az_1-3 a z_2-2 a+b_2 z_1-b_1 z_2+2 b_1 z_1 z_2+2 b_2 z_1 z_2-b_1-z_1-z_2-1\right)
\theta _1
\nn\\
&&+\left(-3 a z_1-a z_2-2 a-b_2 z_1+b_1 z_2+2 b_1 z_1 z_2
\right.
\nn\\
&&
\left. 
+2 b_2 z_1 z_2-b_2-z_1-z_2-1\right)
\theta _2)
G_1(\vec{\gamma};z_1,z_2).
\end{eqnarray}

\begin{eqnarray}
A_{a,G_1}
&=&
\frac{\left(a+b_2\right) \left(a-b_2 z_1+b_1\right)+z_2 \left(2 (a+1) z_1 \left(2 a+b_1+b_2\right)-b_1 \left(a+b_1\right)\right)}{\left(a+b_1\right) \left(a+b_2\right)},
\\
B_{a,G_1}
&=&
\frac{z_2 \left(2 z_1 \left(2 a+b_1+b_2\right)+a+b_1\right)-\left(z_1+1\right) \left(a+b_2\right)}{\left(a+b_1\right) \left(a+b_2\right)},
\\
C_{a,G_1}
&=&
\frac{z_1 \left(2 z_2 \left(2 a+b_2\right)+a+b_2\right)-a \left(z_2+1\right)+b_1 \left(\left(2 z_1-1\right) z_2-1\right)}{\left(a+b_1\right) \left(a+b_2\right)},
\\
D_{a,G_1}
&=&
0,
\end{eqnarray}

\begin{eqnarray}
A_{b_1,G_1}
&=&
\frac{b_1 \left(-2 a z_2+a+b_1+b_2\right)}{\left(b_1+b_2\right) \left(a+b_1\right)},
\\
B_{b_1,G_1}
&=&
\frac{b_1 \left(z_1 \left(1-2 z_2\right)+1\right)}{\left(b_1+b_2\right) z_1 \left(a+b_1\right)},
\\
C_{b_1,G_1}
&=&
-\frac{b_1 \left(2 z_2+1\right)}{\left(b_1+b_2\right) \left(a+b_1\right)},
\\
D_{b_1,G_1}
&=&
0,
\end{eqnarray}

\begin{eqnarray}
A_{b_2,G_1}
&=&
A_{b_1,G_1}(z_1\leftrightarrow z_2,b_1 \leftrightarrow b_2),
\\
B_{b_2,G_1}
&=&
C_{b_1,G_1}(z_1\leftrightarrow z_2,b_1 \leftrightarrow b_2),
\\
C_{b_2,G_1}
&=&
B_{b_1,G_1}(z_1\leftrightarrow z_2,b_1 \leftrightarrow b_2),
\\
D_{b_2,G_1}
&=&
0.
\end{eqnarray}

\boldmath
\subsubsection{Function $\left.G_2(a_1,a_2,b_1,b_2,z_1,z_2\right)$}
\unboldmath

\begin{eqnarray}
G_2\left(a_1,a_2,b_1,b_2,z_1,z_2\right)
=\sum_{n_1,n_2}
\frac{\left(a_1\right)_{n_1} \left(a_2\right)_{n_2} \left(b_1\right)_{n_2-n_1} \left(b_2\right)_{n_1-n_2}}{n_1! n_2!}
z_1^{n_1} z_2^{n_2}.
\end{eqnarray}

The additional PDE reads:
\begin{eqnarray}
\theta _1 \theta _2G_2(\vec{\gamma};\vec{\sigma};z_1,z_2)
=
(-\frac{a_1 a_2 z_1 z_2}{z_1 z_2-1}
-\frac{a_2 z_1 z_2}{z_1 z_2-1}
\theta _1
-\frac{a_1 z_1 z_2}{z_1 z_2-1}
\theta _2)G_2(\vec{\gamma};\vec{\sigma};z_1,z_2).
\end{eqnarray}

\begin{eqnarray}
A_{a_1,G_2}
&=&
\frac{z_1 \left(a_2 z_2-b_2\right)}{a_1+b_1}+1,
\\
B_{a_1,G_2}
&=&
-\frac{z_1+1}{a_1+b_1},
\\
C_{a_1,G_2}
&=&
\frac{z_1 \left(z_2+1\right)}{a_1+b_1},
\\
D_{a_1,G_2}
&=&
0,
\end{eqnarray}

\begin{eqnarray}
A_{a_2,G_2}
&=&
A_{a_1,G_2}(z_1\leftrightarrow z_2,a_1 \leftrightarrow a_2,b_1 \leftrightarrow b_2),
\\
B_{a_2,G_2}
&=&
C_{a_1,G_2}(z_1\leftrightarrow z_2,a_1 \leftrightarrow a_2,b_1 \leftrightarrow b_2),
\\
C_{a_2,G_2}
&=&
B_{a_1,G_2}(z_1\leftrightarrow z_2,a_1 \leftrightarrow a_2,b_1 \leftrightarrow b_2),
\\
D_{a_2,G_2}
&=&
0,
\end{eqnarray}

\begin{eqnarray}
A_{b_1,G_2}
&=&
\frac{b_1 \left(-a_2 z_2+a_1+b_1+b_2\right)}{\left(b_1+b_2\right) \left(a_1+b_1\right)},
\\
B_{b_1,G_2}
&=&
\frac{b_1 \left(z_1+1\right)}{\left(b_1+b_2\right) z_1 \left(a_1+b_1\right)},
\\
C_{b_1,G_2}
&=&
-\frac{b_1 \left(z_2+1\right)}{\left(b_1+b_2\right) \left(a_1+b_1\right)},
\\
D_{b_1,G_2}
&=&
0,
\end{eqnarray}

\begin{eqnarray}
A_{b_2,G_2}
&=&
A_{b_1,G_2}(z_1\leftrightarrow z_2,a_1 \leftrightarrow a_2,b_1 \leftrightarrow b_2),
\\
B_{b_2,G_2}
&=&
C_{b_1,G_2}(z_1\leftrightarrow z_2,a_1 \leftrightarrow a_2,b_1 \leftrightarrow b_2),
\\
C_{b_2,G_2}
&=&
B_{b_1,G_2}(z_1\leftrightarrow z_2,a_1 \leftrightarrow a_2,b_1 \leftrightarrow b_2),
\\
D_{b_2,G_2}
&=&
0.
\end{eqnarray}

\boldmath
\subsubsection{Function $\left.G_3(a_1,a_2,z_1,z_2\right)$}
\unboldmath

\begin{eqnarray}
G_3\left(a_1,a_2,z_1,z_2\right)
=\sum_{n_1,n_2}
\frac{\left(a_1\right)_{2 n_2-n_1} \left(a_2\right)_{2 n_1-n_2}}{n_1! n_2!}
z_1^{n_1} z_2^{n_2}.
\end{eqnarray}

The additional PDE reads:
\begin{eqnarray}
(&-&a_1 a_2 z_1 z_2
+(
a_2 z_1 z_2-2 a_1 z_1 z_2
\left)\theta _1\right.
+(
2 z_1 z_2
\left)\theta _1{}^2\right.
\nn
\\
&+&(
a_1 z_1 z_2-2 a_2 z_1 z_2
\left)\theta _2\right.
+(
2 z_1 z_2
\left)\theta _2{}^2\right.
+(
1-5 z_1 z_2
\left)\theta _1\theta _2\right.)
G_3(\vec{\gamma};z_1,z_2)
=0.
\end{eqnarray}

\begin{eqnarray}
A_{a_1,G_3}
&=&
\frac{2 a_1 \left(2 a_1+2 a_2+1\right)}{\left(2 a_1+a_2\right) \left(2 a_1+a_2+1\right)}
\nn
\\
&&-\frac{3 a_1 z_2 \left(\left(a_1+1\right) \left(5 a_1+4 a_2+2\right)-3 a_2 \left(2 a_1+a_2+1\right) z_1\right)}{\left(2 a_1+a_2\right) \left(2 a_1+a_2+1\right)
   \left(a_1+2 a_2\right)},
\\
B_{a_1,G_3}
&=&
\frac{a_1 \left(\left(a_1+2 a_2\right) \left(4 z_1+1\right)+3 z_1 z_2 \left(6 \left(2 a_1+a_2+1\right) z_1+5 a_1+4 a_2+2\right)\right)}{\left(2 a_1+a_2\right) \left(2 a_1+a_2+1\right) \left(a_1+2 a_2\right) z_1},
\\
C_{a_1,G_3}
&=&
\frac{a_1 \left(-3 z_2 \left(3 \left(2 a_1+a_2+1\right) z_1+10 a_1+8 a_2+4\right)-8 a_1-7 a_2-3\right)}{\left(2 a_1+a_2\right) \left(2 a_1+a_2+1\right) \left(a_1+2 a_2\right)},
\\
D_{a_1,G_3}
&=&
0,
\end{eqnarray}

\begin{eqnarray}
A_{a_2,G_3}
&=&
A_{a_1,G_3}(z_1\leftrightarrow z_2,a_1 \leftrightarrow a_2),
\\
B_{a_2,G_3}
&=&
B_{a_1,G_3}(z_1\leftrightarrow z_2,a_1 \leftrightarrow a_2),
\\
C_{a_2,G_3}
&=&
C_{a_1,G_3}(z_1\leftrightarrow z_2,a_1 \leftrightarrow a_2),
\\
D_{a_2,G_3}
&=&
0.
\end{eqnarray}

\boldmath
\subsubsection{Function $\left.H_1(a,b,c,d,z_1,z_2\right)$}
\unboldmath

\begin{eqnarray}
\left.H_1\text{(a,b,c,d,}z_1,z_2\right)
=\sum_{n_1,n_2}
\frac{(a)_{n_1-n_2} (b)_{n_1+n_2} (c)_{n_2}}{n_1! n_2! (d)_{n_1}}
z_1^{n_1} z_2^{n_2},
\end{eqnarray}

\begin{eqnarray}
A_{a,H_1}
&=&
\frac{a b z_1 \left(a^2+2 a (b+3 c-d+1)+b^2+2 b (c-d+1)+4 c (c-d+1)\right)}{\left(z_2+1\right) (a+b) (a+c) (a+b-2 d+2) (a+c-d+1)}
\nn
\\
&&+\frac{a b z_1 z_2}{\left(z_2+1\right) (a+c) (a+c-d+1)}+\frac{a (a+b+c)}{(a+b) (a+c)},
\\
B_{a,H_1}
&=&
-\frac{a \left(a^2-2 d (a+b+c)+2 a b+3 a c+2 a+b^2+b c+2 (b+c)+2 c^2\right)}{(a+b) (a+c) (a+b-2 d+2) (a+c-d+1)}
\nn\\
&&+\frac{a z_1 \left(a^2+2 a (b+3 c-d+1)+b^2+2 b (c-d+1)+4 c (c-d+1)\right)}{\left(z_2+1\right) (a+b) (a+c) (a+b-2 d+2) (a+c-d+1)}
\nn\\
&&+\frac{a z_1z_2}{\left(z_2+1\right) (a+c) (a+c-d+1)},
\\
C_{a,H_1}
&=&
\frac{a z_1 \left(7 a^2+2 a (5 b+5 c-3 d+6)+3 b^2+b (6 c-6 d+8)+4 (c+1) (c-d+1)\right)}{\left(z_2+1\right) (a+b) (a+c) (a+b-2 d+2) (a+c-d+1)}
\nn
\\
&&
+\frac{a \left(z_2+1\right)}{z_2 (a+b) (a+c)}
+\frac{a z_2 z_1}{\left(z_2+1\right) (a+c) (a+c-d+1)},
\\
D_{a,H_1}
&=&
-\frac{a \left(z_2 \left(-4 z_1+z_2+2\right)+1\right) (3 a+b+2 c-2 d+2)}{z_2 \left(z_2+1\right) (a+b) (a+c) (a+b-2 d+2) (a+c-d+1)},
\end{eqnarray}

\begin{eqnarray}
A_{b,H_1}
&=&
\frac{z_1 z_2 \left(a^3+a (b (b-2 (c+d-1))-2 c)-2 (b+1) c (3 b-2 d+2)\right)}{\left(z_2+1\right) (a+b) (b-d+1) (a+b-2 d+2)}
\nn\\
&&
+\frac{2 a^2 z_1 z_2}{\left(z_2+1\right) (a+b) (a+b-2 d+2)}
-\frac{c z_2}{a+b}+\frac{a z_1}{\left(z_2+1\right) (b-d+1)}
+1,
\\
B_{b,H_1}
&=&
\frac{z_2 z_1 \left(a^2+2 a (b-c-d+1)+b^2-2 b (3 c+d-1)+4 c (d-1)\right)}{\left(z_2+1\right) (a+b) (b-d+1) (a+b-2 d+2)}
\nn\\
&&
+\frac{c z_2 (a+3 b-2 d+2)}{(a+b) (b-d+1) (a+b-2 d+2)}
+\frac{z_1}{\left(z_2+1\right) (b-d+1)}-\frac{1}{b-d+1},
\\
C_{b,H_1}
&=&
-\frac{z_2 z_1 \left(3 a^2+2 a (5 b+c-3 d+4)+7 b^2+6 b (c-d+2)-4 (c+1) (d-1)\right)}{\left(z_2+1\right) (a+b) (b-d+1) (a+b-2 d+2)}
\nn\\
&&
-\frac{z_2+1}{a+b}-\frac{z_1}{\left(z_2+1\right) (b-d+1)},
\\
D_{b,H_1}
&=&
\frac{\left(z_2 \left(-4 z_1+z_2+2\right)+1\right) (a+3 b-2 d+2)}{\left(z_2+1\right) (a+b) (b-d+1) (a+b-2 d+2)},
\end{eqnarray}

\begin{eqnarray}
A_{c,H_1}
&=&
\frac{z_2 \left((a+c-d+1) \left(a-b z_2-b+c\right)+b z_1 \left(z_2 (a-b+2 c+1)+a-b-1\right)\right)}{\left(z_2+1\right) (a+c) (a+c-d+1)}
\nn\\
&&+\frac{1}{z_2+1},
\\
B_{c,H_1}
&=&
\frac{z_2 \left(z_1 \left(z_2 (a-b+2 c+1)+a-b-1\right)-\left(z_2+1\right) (a-b+c)\right)}{\left(z_2+1\right) (a+c) (a+c-d+1)},
\\
C_{c,H_1}
&=&
\frac{z_1 z_2 \left(z_2 (a-b+2 c+1)-a-3 b-1\right)}{\left(z_2+1\right) (a+c) (a+c-d+1)}-\frac{z_2+1}{a+c},
\\
D_{c,H_1}
&=&
\frac{z_2 \left(-4 z_1+z_2+2\right)+1}{\left(z_2+1\right) (a+c) (a+c-d+1)},
\end{eqnarray}

\begin{eqnarray}
A_{d,H_1}
&=&
-\frac{(d-1) \left(-4 d (b+c)+b (b+4 c+5)+6 c+4 d^2-10 d+6\right)}{\left(z_2+1\right) (a+b-2 d+2) (a+b-2 d+3) (a+c-d+1)}
\nn\\
&&
-\frac{(d-1)}{\left(z_2+1\right) (b-d+1) (a+b-2 d+2) (a+b-2 d+3) (a+c-d+1)}
\nn\\
&&
\times
\left(a^3+a^2 (3 b+c-5 d+6)+z_2 \left(a^3+a^2 (3 b+c-5 d+6)+a \left(-2 d (5 b+2 c)    \right.\right.\right.
\nn\\
&&
 \left.       +b (3 b+c+12)+5 c+8 d^2-19 d+11\right)+b^3+b^2 (2 c-5 d+6)
\nn\\
&&
  \left. \left.     +b \left(-4 c d-2 (c-3) c+8 d^2-19 d+11\right)+2 (d-1) (2 d-3) (c-d+1)\right)    \right.
\nn\\
&&
 \left. +a \left(-2 d (5 b+2 c)+3 b (b+c+4)+5 c+8 d^2-19 d+11\right)\right),
\\
B_{d,H_1}
&=&
-\frac{(d-1)}{\left(z_2+1\right) (b-d+1) (a+b-2 d+2) (a+b-2 d+3) (a+c-d+1)}
\nn\\
&&
\times
\left(z_2 \left(a^2+a (2 b-c-4 d+5)+b^2+b (c-4 d+5)-2 c^2+c+4 d^2-10 d+6\right) \right.
\nn\\
&&
\left. +a^2+a (2 b+c-4 d+5)-4 d (b+c)+b (b+3 c+5)+5 c+4 d^2-10 d+6\right)
\nn\\
&&
+\frac{(d-1) \left(a^2+a (2 b+c-4 d+5)-d (4 b+3 c+10)+(b+2) (b+2 c+3)+4 d^2\right)}{z_1 (b-d+1) (a+b-2 d+2) (a+b-2 d+3) (a+c-d+1)}
\nn\\
&&
-\frac{c (d-1) z_2}{z_1 (b-d+1) (a+b-2 d+2) (a+b-2 d+3)},
\\
C_{d,H_1}
&=&
\frac{(d-1) z_2 \left(3 a^2+a (2 b+5 c-4 d+5)-b^2+b (3 c-1)+c (2 c-4 d+5)\right)}{\left(z_2+1\right) (b-d+1) (a+b-2 d+2) (a+b-2 d+3) (a+c-d+1)}
\nn\\
&&
+\frac{(d-1) \left(a^2+a (-2 b+c+1)-b (3 b+c-4 d+5)+c\right)}{\left(z_2+1\right) (b-d+1) (a+b-2 d+2) (a+b-2
   d+3) (a+c-d+1)},
\\
D_{d,H_1}
&=&
\frac{(d-1) \left(z_2 \left(-4 z_1+z_2+2\right)+1\right) \left(-z_2 (a+c-d+1)+b-d+1\right)}{z_1 z_2 \left(z_2+1\right) (b-d+1) (a+b-2 d+2) (a+b-2 d+3) (a+c-d+1)}.
\end{eqnarray}

\boldmath
\subsubsection{Function $\left.H_2(a,b,c,d,e,z_1,z_2\right)$}
\unboldmath

\begin{eqnarray}
\left.H_2\text{(a,b,c,d,e,}z_1,z_2\right)
=\sum_{n_1,n_2}
\frac{(a)_{n_1-n_2} (b)_{n_1} (c)_{n_2} (d)_{n_2}}{n_1! n_2! (e)_{n_1}}
z_1^{n_1} z_2^{n_2},
\end{eqnarray}

\begin{eqnarray}
A_{a,H_2}
&=&
\frac{a b z_1 \left(a^2-e (a+c+d)+2 a c+2 a d+a+c^2+c d+c+d^2+d\right)}{(a+c) (a+d) (a+c-e+1) (a+d-e+1)}
\nn\\
&&
+\frac{a (a+c+d)}{(a+c) (a+d)},
\\
B_{a,H_2}
&=&
\frac{a \left(z_1-1\right) \left(a^2-e (a+c+d)+2 a c+2 a d+a+c^2+c d+c+d^2+d\right)}{(a+c) (a+d) (a+c-e+1) (a+d-e+1)},
\\
C_{a,H_2}
&=&
\frac{a b z_1 (2 a+c+d-e+1)}{(a+c) (a+d) (a+c-e+1) (a+d-e+1)}
\nn\\
&&+\frac{a \left(z_2+1\right)}{z_2 (a+c) (a+d)},
\\
D_{a,H_2}
&=&
\frac{a \left(\left(z_1-1\right) z_2-1\right) (2 a+c+d-e+1)}{z_2 (a+c) (a+d) (a+c-e+1) (a+d-e+1)},
\end{eqnarray}

\begin{eqnarray}
A_{b,H_2}
&=&
\frac{a z_1}{b-e+1}+1,
\\
B_{b,H_2}
&=&
\frac{z_1-1}{b-e+1},
\\
C_{b,H_2}
&=&
-\frac{z_1}{b-e+1},
\\
D_{b,H_2}
&=&
0,
\end{eqnarray}

\begin{eqnarray}
A_{c,H_2}
&=&
1-\frac{d z_2 \left(a+b z_1+c-e+1\right)}{(a+c) (a+c-e+1)},
\\
B_{c,H_2}
&=&
-\frac{d \left(z_1-1\right) z_2}{(a+c) (a+c-e+1)},
\\
C_{c,H_2}
&=&
-\frac{z_2 \left(a+b z_1+c-e+1\right)+a+c-e+1}{(a+c) (a+c-e+1)},
\\
D_{c,H_2}
&=&
\frac{1-\left(z_1-1\right) z_2}{(a+c) (a+c-e+1)},
\end{eqnarray}

\begin{eqnarray}
A_{d,H_2}
&=&
A_{c,H_2}(c \leftrightarrow d),
\\
B_{d,H_2}
&=&
B_{c,H_2}(c \leftrightarrow d),
\\
C_{d,H_2}
&=&
C_{c,H_2}(c \leftrightarrow d),
\\
D_{d,H_2}
&=&
D_{c,H_2}(c \leftrightarrow d),
\end{eqnarray}

\begin{eqnarray}
A_{e,H_2}
&=&
-\frac{(e-1)}{(b-e+1) (a+c-e+1) (a+d-e+1)}
\nn\\
&&
\times
 \left(a^2+a (b+c+d-2 e+2)+b (c+d-e+1)+(c-e+1) (d-e+1)\right),
\\
B_{e,H_2}
&=&
-\frac{(e-1) \left(z_1-1\right) (a+c+d-e+1)}{z_1 (b-e+1) (a+c-e+1) (a+d-e+1)},
\\
C_{e,H_2}
&=&
\frac{b-b e}{(b-e+1) (a+c-e+1) (a+d-e+1)},
\\
D_{e,H_2}
&=&
-\frac{(e-1) \left(\left(z_1-1\right) z_2-1\right)}{z_1 z_2 (b-e+1) (a+c-e+1) (a+d-e+1)}.
\end{eqnarray}

\boldmath
\subsubsection{Function $\left.H_3(a,b,c,z_1,z_2\right)$}
\unboldmath

\begin{eqnarray}
H_3\left(a,b,c,z_1,z_2\right)
=\sum_{n_1,n_2}
\frac{(a)_{2 n_1+n_2} (b)_{n_2}}{n_1! n_2! (c)_{n_1+n_2}}
z_1^{n_1} z_2^{n_2}.
\end{eqnarray}

The additional PDE reads:
\begin{eqnarray}
(b z_2 \theta _1
-a z_1\theta _2
-z_1\theta _2^2
+(z_2-2 z_1)\theta _1\theta _2)
H_3(\vec{\gamma};\vec{\sigma};z_1,z_2)
=0.
\end{eqnarray}

\begin{eqnarray}
A_{a,H_3}
&=&
\frac{2 (a+1) z_1 (2 a+b-2 c+2)+(a+b-2 c+2) \left(a+b z_2-c+1\right)}{(a-c+1) (a+b-2 c+2)},
\\
B_{a,H_3}
&=&
\frac{\left(4 z_1-1\right) (2 a+b-2 c+2)}{(a-c+1) (a+b-2 c+2)},
\\
C_{a,H_3}
&=&
\frac{\left(z_2-1\right) z_2 (a+b-2 c+2)+z_1 \left(2 z_2 (2 a+b-2 c+2)-a\right)}{z_2 (a-c+1) (a+b-2 c+2)},
\\
D_{a,H_3}
&=&
0,
\end{eqnarray}

\begin{eqnarray}
A_{b,H_3}
&=&
\frac{2 a z_2}{a+b-2 c+2}+1,
\\
B_{b,H_3}
&=&
\frac{\left(4 z_1-1\right) z_2}{z_1 (a+b-2 c+2)},
\\
C_{b,H_3}
&=&
\frac{2 z_2-1}{a+b-2 c+2},
\\
D_{b,H_3}
&=&
0,
\end{eqnarray}

\begin{eqnarray}
A_{c,H_3}
&=&
\frac{(c-1) \left(-z_2 \left(4 a^2+a (3 b-8 c+10)+(b-2 c+2) (b-2 c+3)\right)-2 a b z_1\right)}{z_2 (a-c+1) (a+b-2 c+2) (a+b-2 c+3)},
\\
B_{c,H_3}
&=&
-\frac{(c-1) \left(4 z_1-1\right) \left(z_2 (a-c+1)+b z_1\right)}{z_1 z_2 (a-c+1) (a+b-2 c+2) (a+b-2 c+3)},
\\
C_{c,H_3}
&=&
\frac{(c-1)}{z_2^2 (a-c+1) (a+b-2 c+2) (a+b-2 c+3)}
\nn\\&&
 \times
\left(z_2 \left(-z_2 (3 a+b-4 c+5)+2 a+b-3 c+4\right)-z_1 \left(a+2 b z_2-2 c+3\right)\right),
\\
D_{c,H_3}
&=&
0.
\end{eqnarray}

\boldmath
\subsubsection{Function $\left.H_4(a,b,c,d,z_1,z_2\right)$}
\unboldmath

\begin{eqnarray}
\left.H_4\text{(a,b,c,d,}z_1,z_2\right)
=\sum_{n_1,n_2}
\frac{(a)_{2 n_1+n_2} (b)_{n_2}}{n_1! n_2! (c)_{n_1} (d)_{n_2}}
z_1^{n_1} z_2^{n_2},
\end{eqnarray}

\begin{eqnarray}
A_{a,H_4}
&=&
\frac{4 (a+1) z_2 z_1 \left(a^2+2 a (b-c-d+2)-(b-d+1) (2 c+d-3)\right)}{\left(z_2-1\right) (a-2 c+2) (a-d+1) (a-2 c-d+3)}
\nn\\
&&+\frac{b z_2}{a-d+1}-\frac{4 (a+1) z_1}{\left(z_2-1\right) (a-2 c+2)}+1,
%
%
\\
B_{a,H_4}
&=&
\frac{2 z_2 \left(4 z_1 \left(a^2+2 a (b-c-d+2)-(b-d+1) (2 c+d-3)\right)-a b \left(z_2-1\right)\right)}{\left(z_2-1\right) (a-2 c+2) (a-d+1) (a-2 c-d+3)}
\nn\\
&&
-\frac{2 b z_2}{(a-2 c+2) (a-d+1)}
-\frac{8 z_1}{\left(z_2-1\right) (a-2 c+2)}
-\frac{2}{a-2 c+2},
\\
C_{a,H_4}
&=&
\frac{4 z_1 \left(z_2 (a+2 b-2 d+2)+a+2 c-d+1\right)}{\left(z_2-1\right) (a-d+1) (a-2 c-d+3)}
\nn
\\ &&
+\frac{4 z_1 (2 c-d-1) \left(z_2 (b-d+1)+2 c-1\right)}{\left(z_2-1\right) (a-2 c+2) (a-d+1) (a-2 c-d+3)}
\nn
\\ &&
+\frac{z_2-1}{a-d+1},
\\
D_{a,H_4}
&=&
-\frac{2 \left(\left(z_2-1\right){}^2-4 z_1\right) (2 a-2 c-d+3)}{\left(z_2-1\right) (a-2 c+2) (a-d+1) (a-2 c-d+3)},
\end{eqnarray}

\begin{eqnarray}
A_{b,H_4}
&=&
\frac{a z_2}{b-d+1}+1,
\\
B_{b,H_4}
&=&
\frac{2 z_2}{b-d+1},
\\
C_{b,H_4}
&=&
\frac{z_2-1}{b-d+1},
\\
D_{b,H_4}
&=&
0,
\end{eqnarray}

\begin{eqnarray}
A_{c,H_4}
&=&
\frac{2 (c-1) z_2}{\left(z_2-1\right) (a-2 c+2) (a-2 c+3) (a-2 c-d+3) (a-2 c-d+4)}
\nn\\
&&
 \times
\left(-2 a^3-a^2 (b-10 c-4 d+17) \right.
\nn\\
&&
\left.
+a (2 (b (c-1)-(2 c+d) (4 c+d))+54 c+20 d-45) \right.
\nn\\
&&
\left.
+a b z_2 (-b+d-1)+(2 c-3) (2 c+d-4) (2 c+d-3)\right)
\nn\\
&&
+\frac{2 (c-1) (2 a-2   c+3)}{\left(z_2-1\right) (a-2 c+2) (a-2 c+3)},
\\
B_{c,H_4}
&=&
-\frac{4 (c-1) z_2 (a+b-2 c-d+4)}{\left(z_2-1\right) (a-2 c+2) (a-2 c+3) (a-2 c-d+4)}
\nn\\
&&
+\frac{b (c-1) z_2 \left(z_2 (-a+b+2 c-3)+3 a-2 (3 c+d-5)\right)}{z_1 (a-2 c+2) (a-2 c+3) (a-2 c-d+3) (a-2 c-d+4)}
\nn\\
&&
-\frac{4 b (c-1) z_2 \left(z_2  (b-d+1)+d-1\right)}{\left(z_2-1\right) (a-2 c+2) (a-2 c+3) (a-2 c-d+3) (a-2 c-d+4)}
\nn\\
&&
+\frac{c-1}{z_1 (a-2 c+2) (a-2 c+3)}
+\frac{4 (c-1)}{\left(z_2-1\right) (a-2 c+2) (a-2 c+3)},
\\
C_{c,H_4}
&=&
\frac{2 (c-1) \left(z_2 (a-2 b+1)+2 d\right)}{\left(z_2-1\right) (a-2 c+2) (a-2 c-d+3) (a-2 c-d+4)}
\nn\\
&&
+\frac{2 (c-1) \left(d-b z_2\right) \left(z_2 (b-d+1)+2 c-3\right)}{\left(z_2-1\right) (a-2 c+2) (a-2 c+3) (a-2 c-d+3) (a-2 c-d+4)}
\nn\\
&&
-\frac{4 (a+1)   (c-1)}{\left(z_2-1\right) (a-2 c+2) (a-2 c-d+3) (a-2 c-d+4)},
\\
D_{c,H_4}
&=&
\frac{(c-1) \left(\left(z_2-1\right){}^2-4 z_1\right) \left(z_2 (-a+b+2 c-3)+2 a-4 c-d+6\right)}{z_1 \left(z_2-1\right) (a-2 c+2) (a-2 c+3) (a-2 c-d+3) (a-2 c-d+4)},
\end{eqnarray}

\begin{eqnarray}
A_{d,H_4}
&=&
\frac{(d-1) \left(-\left(z_2-1\right) (a+b-d+1) (a-2 c-d+3)-4 a b z_1\right)}{\left(z_2-1\right) (a-d+1) (b-d+1) (a-2 c-d+3)},
\\
B_{d,H_4}
&=&
\frac{2 b (d-1) \left(4 z_1-z_2+1\right)}{\left(z_2-1\right) (a-d+1) (-b+d-1) (a-2 c-d+3)},
\\
C_{d,H_4}
&=&
\frac{(d-1) \left(4 z_1 \left(a+b z_2-d+2\right)+\left(z_2-1\right){}^2 (a-2 c-d+3)\right)}{\left(z_2-1\right) z_2 (a-d+1) (-b+d-1) (a-2 c-d+3)},
\\
D_{d,H_4}
&=&
-\frac{2 (d-1) \left(\left(z_2-1\right){}^2-4 z_1\right)}{\left(z_2-1\right) z_2 (a-d+1) (-b+d-1) (a-2 c-d+3)}.
\end{eqnarray}

\boldmath
\subsubsection{Function $\left.H_5(a,b,c,z_1,z_2\right)$}
\unboldmath

\begin{eqnarray}
\left.H_5\text{(a,b,c,}z_1,z_2\right)
=\sum_{n_1,n_2}
\frac{(a)_{2 n_1+n_2} (b)_{n_2-n_1}}{n_1! n_2! (c)_{n_2}}
z_1^{n_1} z_2^{n_2},
\end{eqnarray}

\begin{eqnarray}
A_{a,H_5}
&=&
\frac{1}{\left(z_2+z_1 \left(6 z_2-4\right)-1\right) (a+2 b) (a-c+1) (a+2 b-3 c+3)}
\nn\\
&&
\times
2 z_1 \left(4 (a+1) z_1 \left(2 (a-c+1) (a+2 b-3 c+3) \right.\right.
\nn\\
&&
\left.\left.
-z_2 \left(7 a^2+a (16 b-15 c+23)+(2 b-3 c+3) (2 b-3 c+5)\right)\right)  \right.
\nn\\
&&
\left.
+z_2 \left(13 a^3+a^2 (4 b-13 c+47) \right. \right.
\nn\\
&&
\left.\left.
-3 z_2 \left(4 a^3-3 a^2 (b+c-5)+a (-2 b (3 b-3 c+2)-9 c+17) \right.\right.\right.
\nn\\
&&
\left.\left. \left.
-\left(2 b^2+b-2\right) (2   b-3 c+3)\right)+a (b (20-14 c)+c (3 c-25)+46) \right.\right.
\nn\\
&&
\left.\left. 
-2 (2 b-3 c+3) (b (2 b+3 c-2)-c-2)\right)\right)
\nn\\
&&
+\frac{\left(z_2-1\right) \left(a+b z_2-c+1\right)}{\left(z_2+z_1 \left(6 z_2-4\right)-1\right) (a-c+1)}
\nn\\
&&
-\frac{4 (2 b-1) z_1}{\left(z_2+z_1 \left(6 z_2-4\right)-1\right) (a+2 b)},
\\
B_{a,H_5}
&=&
-\frac{2 z_1 z_2 }{\left(z_2+z_1 \left(6 z_2-4\right)-1\right) (a+2 b) (a-c+1) (a+2 b-3 c+3)}
\nn\\
&&
\times
\left(8 z_1 \left(7 a^2+a (16 b-15 c+23)+(2 b-3 c+3) (2 b-3 c+5)\right) \right. 
\nn\\
&&
\left.
+3 z_2 \left(21 a^2+2 a (15 b-9 c+23)+(6 b+7) (2 b-3 c+3)\right)-48 a^2 \right.
\nn\\
&&
\left.
+8 a (-8 b+c-11)-2 (2 b-3 c+3) (8 b+5 c+5)\right)
\nn\\
&&
-\frac{\left(z_2-1\right) \left(z_2 (a+2 b)+2 (a-c+1)\right)}{\left(z_2+z_1 \left(6 z_2-4\right)-1\right) (a+2 b) (a-c+1)}
\nn\\
&&
+\frac{16 z_1 \left(2 z_1+1\right)}{\left(z_2+z_1 \left(6 z_2-4\right)-1\right) (a+2 b)},
\\
C_{a,H_5}
&=&
\frac{1}{\left(z_2+z_1 \left(6 z_2-4\right)-1\right) (a+2 b) (a-c+1) (a+2 b-3 c+3)}
\nn\\
&&
\times
(-8 z_1^2 \left(z_2 \left(7 a^2+a (16 b-15 c+23)+(2 b-3 c+3) (2 b-3 c+5)\right)  \right.
\nn\\
&&
\left.
+2 a (3 a-3 c+7)+4 b-6 c+6\right)
\nn\\
&&
-2 z_1 \left(z_2 \left(3 z_2 \left(3 a^2+a (8-6 b)-(3 b-2) (2 b-3 c+3)\right)-17 a^2 \right.\right.
\nn\\
&&
\left.\left.
+a (20 b+7 c-43)+(2 b-3 c+3) (12 b+c-10)\right)+4   a^2-8 a (b-1) \right.
\nn\\
&&
\left.
-2 (2 b-1) (2 b-3 c+3)\right))
\nn\\
&&
+\frac{\left(z_2-1\right){}^2}{\left(z_2+z_1 \left(6 z_2-4\right)-1\right) (a-c+1)},
\\
D_{a,H_5}
&=&
-\frac{2 \left(-z_2+z_1 \left(16 z_1+9 z_2 \left(3 z_2-4\right)+8\right)+1\right) (4 a+2 b-3 c+3)}{\left(z_2+z_1 \left(6 z_2-4\right)-1\right) (a+2 b) (a-c+1) (a+2 b-3 c+3)},
\end{eqnarray}

\begin{eqnarray}
A_{b,H_5}
&=&
\frac{b}{\left(z_2+z_1 \left(6 z_2-4\right)-1\right) (a+2 b) (a+2 b+1) (a+2 b-3 c+3)(a+2 b-3 c+4)}
\nn\\
&&
\times
\left(2 z_1 \left(6 z_2 \left(a^3+2 a^2 (3 b-3 c+5)+a \left(12 b^2-24 (b+2) c+36 b+18 c^2+31\right) \right.\right.\right.
\nn\\
&&
\left.\left.\left.
+(2 b+1) (2 b-3 c+3) (2 b-3 c+4)\right)+27 a z_2^2 (a+2 b+1) (a+2 b-3 c+4)  \right.\right.
\nn\\
&&
\left.\left.
-4 (2 a+2 b+1) (a+2 b-3 c+3) (a+2 b-3 c+4)\right)  \right.
\nn\\
&&
\left.
+z_2 \left(-23 a^3+9 a
   z_2 \left(3 a^2-3 c (2 a+b+1)+6 a b+9 a+6 b+4\right)   \right. \right.
\nn\\
&&
\left.\left.
+a^2 (-40 b+30 c-57)+2 a \left(10 b^2+b (7-18 c)+6 c (3 c-5)+7\right)  \right. \right.
\nn\\
&&
\left.\left.
+2 (2 b+1) (2 b-3 c+3) (2 b-3 c+4)\right)\right)
\nn\\
&&
-\frac{2 b (2 a+2 b+1)}{\left(z_2+z_1 \left(6 z_2-4\right)-1\right) (a+2 b) (a+2 b+1)},
\\
B_{b,H_5}
&=&
\frac{-3 b (3 c-2) z_2 \left(3 z_2 (6 b-15 c+7)-2 (8 b-21 c+9)-24 (c-1) z_1\right)}{\left(z_2+z_1 \left(6 z_2-4\right)-1\right) (a+2 b) (a+2 b+1) (a+2 b-3 c+3) (a+2 b-3 c+4)}
\nn\\
&&
+\frac{108 b z_1 z_2^2}{\left(z_2+z_1 \left(6 z_2-4\right)-1\right) (a+2 b) (a+2 b-3 c+3)}
\nn\\
&&
+\frac{b z_2 \left(27 z_2 (4 a+4 b+5 c+2)-4 (23 a+22 b+39 c+5)+48 z_1\right)}{\left(z_2+z_1 \left(6 z_2-4\right)-1\right) (a+2 b) (a+2 b+1) (a+2 b-3 c+4)}
\nn\\
&&
-\frac{16 b z_1^2+8 b z_1}{z_1 \left(z_2+z_1 \left(6 z_2-4\right)-1\right) (a+2 b) (a+2   b+1)}
\nn\\
&&
+\frac{b \left(z_2-1\right)}{z_1 \left(z_2+z_1 \left(6 z_2-4\right)-1\right) (a+2 b) (a+2 b+1)},
\\
C_{b,H_5}
&=&
\frac{b}{\left(z_2+z_1 \left(6 z_2-4\right)-1\right) (a+2 b) (a+2 b+1) (a+2 b-3 c+3) (a+2 b-3 c+4)}
\nn\\
&&
\times
\left(-z_2 \left(12 z_1 \left(3 a^2+a (12 b-9 c+13)+6 (2-3 b) c+2 b (6 b+13)-9 c^2\right)+49 a^2  \right.\right.
\nn\\
&&
\left.\left.
-3 c (35 a+28 b+25)+112 a b+157 a+28 b^2+146 b+9 c^2+84\right) \right.
\nn\\
&&
\left.
+9 z_2^2 \left(3 a^2-3 c (2 a+b+1)+6 a b+9 a+6 b+4\right) \right.
\nn\\
&&
\left.
+2 \left(4 z_1+1\right) \left(8 a^2+2 a (10 b-9 c+13)+(2 b+1) (4 b-9 c+12)\right)\right)
\nn\\
&&
+\frac{54 b z_1 z_2^2}{\left(z_2+z_1 \left(6 z_2-4\right)-1\right) (a+2 b) (a+2 b-3 c+3)},
\\
D_{b,H_5}
&=&
\frac{3 b \left(-z_2+z_1 \left(16 z_1+9 z_2 \left(3 z_2-4\right)+8\right)+1\right) (2 a+4 b-3 c+4)}{z_1 (a+2 b) (a+2 b+1) (a+2 b-3 c+3) (a+2 b-3 c+4)}
\nn\\
&&
\times
\frac{1}{\left(z_2+z_1 \left(6 z_2-4\right)-1\right) },
\end{eqnarray}

\begin{eqnarray}
A_{c,H_5}
&=&
-\frac{a (c-1) }{ (a-c+1) (a+2 b-3 c+3) (a+2 b-3 c+4) (a+2 b-3 c+5)}
\nn\\
&&
\times
\frac{1}{\left(z_2+z_1 \left(6 z_2-4\right)-1\right)}
\left(z_2 \left(9 a^2-9 c (4 a+7 b+17)+17 a b+54 a+20 b^2  \right.\right.
\nn\\
&&
\left.\left.
+87 b+54 c^2+105\right)+2 z_1 \left(3 z_2 \left(4 a^2+a (18 b-27 c+38)+6 (2 b-3 c)^2 \right.\right.\right.
\nn\\
&&
\left.\left.\left.
+94 b-144 c+94\right)-2 \left(2 a^2+a (14 b-23 c+31)+24 b^2-70 (b+2) c \right.\right.\right.
\nn\\
&&
\left.\left.\left.
+90 b+54
   c^2+89\right)\right)-9 a^2-18 a (b-2 c+3)+64 b c-4 b (5 b+22) \right.
\nn\\
&&
\left.
-54 c^2+3 (51 c-35)\right)
\nn\\
&&
+\frac{16 a (c-1) z_1^2 (a+2 b+1)}{\left(z_2+z_1 \left(6   z_2-4\right)-1\right) (a-c+1) (a+2 b-3 c+3) (a+2 b-3 c+5)}
\nn\\
&&
-\frac{(c-1) (2 b-3 c+3) (2 b-3 c+4) (2 b-3 c+5)}{(a-c+1) (a+2 b-3 c+3) (a+2 b-3 c+4) (a+2 b-3 c+5)},
\\
B_{c,H_5}
&=&
\frac{(c-1) }
{(a-c+1) (a+2 b-3 c+3) (a+2 b-3 c+4) (a+2 b-3 c+5)}
\nn\\
&&
\times
\frac{1}{\left(z_2+z_1 \left(6 z_2-4\right)-1\right) }
\left(2 \left(9 a^2+4 z_1 \left(4 z_1 (a+2 b+1) (a+2 b-3 c+4)  \right.\right.\right.
\nn\\
&&
\left.\left.\left.
+c (16 a+26 b+13)-2 (a+2 b+1) (3 a+4 b+9)\right)+a (8 b-13 c+19) \right.\right.
\nn\\
&&
\left.\left.
-2 (b+1) (2 b+1)\right)+z_2 \left(12 z_1 \left(3 a^2+a (12 b-9 c+13) \right.\right.\right.
\nn\\
&&
\left.\left.\left.
+(2 b+1) (6 b-9 c+13)\right)-19 a^2 \right.\right.
\nn\\
&&
\left.\left.
+a (-16 b+27 c-39)+4 (b+1) (2 b+1)\right)\right),
\\
C_{c,H_5}
&=&
\frac{(c-1)}
{ (a-c+1) (a+2 b-3 c+3) (a+2 b-3 c+4) (a+2 b-3 c+5)}
\nn\\
&&
\times
\frac{1}{z_2 \left(z_2+z_1 \left(6 z_2-4\right)-1\right)}
\left(z_2^2 \left(-\left(6 z_1 \left(3 a^2+2 a (6 b-9 c+13)   \right.\right.\right.\right.
\nn\\
&&
\left.\left.\left.\left.
+3 (2 b-3 c)^2+46 b-72 c+47\right)+8 a^2-27 c (a+b+3)+11 a b+42 a  \right.\right.\right.
\nn\\
&&
\left.\left.\left.
+8 b^2+39 b+27 c^2+58\right)\right) \right.
\nn\\
&&
\left.
+z_2 \left(4 z_1 \left(4 a^2+4 z_1 (a+2 b+1) (a+2 b-3 c+4)  \right.\right.\right.
\nn\\
&&
\left.\left.\left.
+a (20 b-32 c+46)-70 b c+6 b (4 b+15)+54 c^2-143 c+93\right)  \right.\right.
\nn\\
&&
\left.\left.
+15 a^2+a (24 b-53 c+81)+2 \left(-28 b c+8 b (b+5)+27 c^2-80 c+57\right)\right)  \right.
\nn\\
&&
\left.
+2 \left(4 z_1+1\right) \left(-3 a^2+4 z_1 (a-c+2) (a+2 b-3 c+4)-6 a (b-2 c+3)  \right.\right.
\nn\\
&&
\left.\left.
+14 b c-4 b (b+5)-13 c^2+38 c-27\right)\right),
\\
D_{c,H_5}
&=&
\frac{(c-1) \left(-z_2+z_1 \left(16 z_1+9 z_2 \left(3 z_2-4\right)+8\right)+1\right)}
{ (a-c+1) (a+2 b-3 c+3) (a+2 b-3 c+4) (a+2 b-3 c+5)}
\nn\\
&&
\times
\frac{ \left(4 z_1 (a+2 b-3 c+4)-a+c-1\right)}{z_1 z_2 \left(z_2+z_1 \left(6 z_2-4\right)-1\right)}.
\end{eqnarray}

\boldmath
\subsubsection{Function $\left.H_6(a,b,c,z_1,z_2\right)$}
\unboldmath

\begin{eqnarray}
H_6\left(a,b,c,z_1,z_2\right)
=\sum_{n_1,n_2}
\frac{(a)_{2 n_1-n_2} (b)_{n_2-n_1} (c)_{n_2}}{n_1! n_2!}
z_1^{n_1} z_2^{n_2}.
\end{eqnarray}

The additional PDE reads:
\begin{eqnarray}
(a c z_1 z_2
+
2 c z_1 z_2
\theta _1
+(
a z_1 z_2-c z_1 z_2
)\theta _2
-z_1 z_2
\theta _2{}^2
+(
2 z_1 z_2-1
)\theta _1\theta _2)
H_6(\vec{\gamma};z_1,z_2)
=0.
\end{eqnarray}

\begin{eqnarray}
A_{a,H_6}
&=&
\frac{a \left(z_1 \left(c z_2 (a+c)-2 (a+1) (2 (a+b)+c)\right)+(a+2 b) (a+b+c)\right)}{(a+b) (a+2 b) (a+c)},
\\
B_{a,H_6}
&=&
-\frac{a \left(4 z_1+1\right) (2 (a+b)+c)}{(a+b) (a+2 b) (a+c)},
\\
C_{a,H_6}
&=&
\frac{a \left(z_2 \left(z_1 \left(4 (a+b)+z_2 (a+c)+2 c\right)+a+2 b\right)+a+2 b\right)}{z_2 (a+b) (a+2 b) (a+c)},
\\
D_{a,H_6}
&=&
0,
\end{eqnarray}

\begin{eqnarray}
A_{b,H_6}
&=&
\frac{b c z_2 \left(z_1 \left(z_2 (a+2 b-c+1)+2 a\right)-2 a-3 b-1\right)}{(a+b) (a+2 b) (a+2 b+1)}
\nn\\
&&
+\frac{2 b (2 a+2 b+1)}{(a+2 b) (a+2 b+1)},
\\
B_{b,H_6}
&=&
\frac{b \left(4 z_1+1\right) \left(a+b+c z_1 z_2\right)}{z_1 (a+b) (a+2 b) (a+2 b+1)},
\\
C_{b,H_6}
&=&
\frac{b \left(z_2 \left(z_1 \left(z_2 (a+2 b-c+1)-2 c\right)-2 a-3 b-1\right)-3 a-4 b-1\right)}{(a+b) (a+2 b) (a+2 b+1)},
\\
D_{b,H_6}
&=&
0,
\end{eqnarray}

\begin{eqnarray}
A_{c,H_6}
&=&
\frac{z_2 \left(2 a z_1-b\right)}{a+c}+1,
\\
B_{c,H_6}
&=&
\frac{\left(4 z_1+1\right) z_2}{a+c},
\\
C_{c,H_6}
&=&
-\frac{2 z_1 z_2+z_2+1}{a+c},
\\
D_{c,H_6}
&=&
0.
\end{eqnarray}

\boldmath
\subsubsection{Function $\left.H_7(a,b,c,d,z_1,z_2\right)$}
\unboldmath

\begin{eqnarray}
\left.H_7\text{(a,b,c,d,}z_1,z_2\right)
=\sum_{n_1,n_2}
\frac{(a)_{2 n_1-n_2} (b)_{n_2} (c)_{n_2}}{n_1! n_2! (d)_{n_1}}
z_1^{n_1} z_2^{n_2},
\end{eqnarray}

\begin{eqnarray}
A_{a,H_7}
&=&
\frac{a (a+b+c)}{(a+b) (a+c)}
+\frac{4 a z_1 }
{\left(z_2+1\right) (a+b)   (a+c) (a+b-2 d+2) (a+c-2 d+2)}
\nn\\
&&
\times
\left((a+1) \left(a^2+2 a (b+c-d+1)+b^2+b (c-2 d+2)+c (c-2 d+2)\right) \right.
\nn\\
&&
\left.
+z_2 \left(a^3+a^2 (2 b+2 c-2 d+3) \right.\right.
\nn\\
&&
\left.\left.
+a \left(b^2-2 d (b+c+1)+3 b c+4 b+c^2+4 c+2\right)+b^2 (c+1)  \right.\right.
\nn\\
&&
\left.\left.
+b (c+1) (c-2 d+2)+c (c-2 d+2)\right)\right),
\\
B_{a,H_7}
&=&
\frac{2 a \left(4 z_1-1\right) \left(a^2+2 a (b+c-d+1)+b^2+(b+c) (c-2 d+2)\right)}{(a+b) (a+c) (a+b-2 d+2) (a+c-2 d+2)},
\\
C_{a,H_7}
&=&
\frac{a}{\left(z_2+1\right) (a+b) (a+c) (a+b-2 d+2) (a+c-2 d+2)}
\nn\\
&&
\times
\left(4 z_1 z_2 (c (a+b)+a (a+b+2)+b+c-2 d+2) \right.
\nn\\
&&
\left.
-4 z_1 \left(a^2+2 a (b+c-d+1)+b^2+b (c-2 d+2)+c (c-2 d+2)\right)\right)
\nn\\
&&
+\frac{a \left(z_2+2\right)}{\left(z_2+1\right) (a+b)(a+c)}
+\frac{a}{\left(z_2^2+z_2\right) (a+b) (a+c)},
\\
D_{a,H_7}
&=&
\frac{2 a \left(z_2 \left(\left(4 z_1-1\right) z_2-2\right)-1\right) (2 a+b+c-2 d+2)}{z_2 \left(z_2+1\right) (a+b) (a+c) (a+b-2 d+2) (a+c-2 d+2)},
\end{eqnarray}

\begin{eqnarray}
A_{b,H_7}
&=&
-\frac{4 c z_1 z_2 \left(z_2 (a+b+1)+a\right)}{\left(z_2+1\right) (a+b) (a+b-2 d+2)}+\frac{z_2 \left(a+b-c z_2-c\right)}{\left(z_2+1\right) (a+b)}+\frac{1}{z_2+1},
\\
B_{b,H_7}
&=&
\frac{2 c \left(1-4 z_1\right) z_2}{(a+b) (a+b-2 d+2)},
\\
C_{b,H_7}
&=&
-\frac{4 z_1 z_2 \left(z_2 (a+b+1)-c\right)+a+b-2 d+2}{\left(z_2+1\right) (a+b) (a+b-2 d+2)}-\frac{z_2 \left(z_2+2\right)}{\left(z_2+1\right) (a+b)},
\\
D_{b,H_7}
&=&
\frac{\left(2-8 z_1\right) z_2^2+4 z_2+2}{\left(z_2+1\right) (a+b) (a+b-2 d+2)},
\end{eqnarray}

\begin{eqnarray}
A_{c,H_7}
&=&
A_{b,H_7}(b\leftrightarrow c),
\\
B_{c,H_7}
&=&
B_{b,H_7}(b\leftrightarrow c),
\\
C_{c,H_7}
&=&
C_{b,H_7}(b\leftrightarrow c),
\\
D_{c,H_7}
&=&
D_{b,H_7}(b\leftrightarrow c),
\end{eqnarray}

\begin{eqnarray}
A_{d,H_7}
&=&
-\frac{2 (d-1) }{z_2 \left(z_2+1\right) (a+b-2 d+2)   (a+b-2 d+3) (a+c-2 d+2) (a+c-2 d+3)}
\nn\\
&&
\times\left(-a b c+z_2 \left(z_2+1\right) \left(2 a^3+a \left(2 \left(b^2-6 d (b+c)+2 b c+8 b+c^2+8 d^2\right) \right.\right.\right.
\nn\\
&&
\left.\left.\left.
+16 c-42 d+27\right)+b^2 (c-2 d+3)+b \left(-8 c d+c (c+11)+8 d^2-22 d+15\right)\right)\right)
\nn\\
&&
-\frac{2 a^2 (d-1) (4 b+4 c-10 d+13)}{(a+b-2 d+2) (a+b-2 d+3) (a+c-2 d+2) (a+c-2 d+3)}
\nn\\
&&
-\frac{2 b c(d-1) z_2 (2 a+b+c-2 d+2)}{\left(z_2+1\right) (a+b-2 d+2) (a+b-2 d+3) (a+c-2 d+2) (a+c-2 d+3)}
\nn\\
&&
+\frac{2 (d-1) (2 d-3) \left(c^2+c (5-4 d)+4 d^2-10 d+6\right)}{(a+b-2 d+2) (a+b-2 d+3) (a+c-2 d+2) (a+c-2 d+3)},
\\
B_{d,H_7}
&=&
-\frac{(d-1) \left(4 z_1-1\right)}{z_1 z_2 (a+b-2 d+2) (a+b-2 d+3) (a+c-2 d+2) (a+c-2 d+3)}
\nn\\
&&
\times\left(z_2 \left(a^2+a (2 b+2 c-4 d+5)+b^2+b (c-4 d+5) \right.\right.
\nn\\
&&
\left.\left.
+(c-2 d+2) (c-2 d+3)\right)-b c\right),
\\
C_{d,H_7}
&=&
\frac{2 (d-1)}{z_2 \left(z_2+1\right) (a+b-2 d+2) (a+b-2 d+3) (a+c-2 d+2) (a+c-2 d+3)}
\nn\\
&&
\times
 \left(-b c-z_2 \left(a^2-2 (a+1) d+4 a-b^2+b c+b-c^2+c+3\right) \right.
\nn\\
&&
\left.
+z_2^2 \left(-\left(2 a^2+2 a (b+c-2 d+3)-2 d (b+c+1)+2 b c+3 b+3 c+3\right)\right)\right),
\\
D_{d,H_7}
&=&
-\frac{(d-1) \left(z_2 \left(\left(4 z_1-1\right) z_2-2\right)-1\right)}
{ (a+b-2 d+3) (a+c-2 d+2) (a+c-2 d+3)}
\nn\\
&&
\times
\frac{ \left(z_2 (2 a+b+c-4 d+5)+a-2 d+3\right)}{z_1 z_2^2 \left(z_2+1\right) (a+b-2 d+2)}.
\end{eqnarray}


\subsection{Confluent Horn-type hypergeometric functions}

\boldmath
\subsubsection{Function $\left.\Phi_1(a,b,c,z_1,z_2\right)$}
\unboldmath

\begin{eqnarray}
\left.\Phi _1\text{(a,b,c,}z_1,z_2\right)
=\sum_{n_1,n_2}
\frac{(a)_{n_1+n_2} (b)_{n_1}}{n_1! n_2! (c)_{n_1+n_2}}
z_1^{n_1} z_2^{n_2}.
\end{eqnarray}

The additional PDE reads:
\begin{eqnarray}
\theta _1 \theta _2\Phi_1(\vec{\gamma};\vec{\sigma};z_1,z_2)
=
(\frac{z_2}{z_1}
\theta _1
-b
\theta _2)
\Phi_1(\vec{\gamma};\vec{\sigma};z_1,z_2).
\end{eqnarray}

\begin{eqnarray}
A_{a,\Phi_1}
&=&
\frac{b z_1+z_2}{a-c+1}+1,
\\
B_{a,\Phi_1}
&=&
\frac{z_1-1}{a-c+1},
\\
C_{a,\Phi_1}
&=&
\frac{1}{-a+c-1},
\\
D_{a,\Phi_1}
&=&
0,
\end{eqnarray}

\begin{eqnarray}
A_{b,\Phi_1}
&=&
1,
\\
B_{b,\Phi_1}
&=&
0,
\\
C_{b,\Phi_1}
&=&
-\frac{z_1}{z_2},
\\
D_{b,\Phi_1}
&=&
0,
\end{eqnarray}

\begin{eqnarray}
A_{c,\Phi_1}
&=&
\frac{a}{-a+c-1}+1,
\\
B_{c,\Phi_1}
&=&
0,
\\
C_{c,\Phi_1}
&=&
\frac{c-1}{z_2 (a-c+1)},
\\
D_{c,\Phi_1}
&=&
0.
\end{eqnarray}

\boldmath
\subsubsection{Function $\left.\Phi_2(b_1,b_2,c,z_1,z_2\right)$}
\unboldmath

\begin{eqnarray}
\Phi _2\left(b_1,b_{2,}\text{c,}z_1,z_2\right)
=\sum_{n_1,n_2}
\frac{\left(b_1\right)_{n_1} \left(b_2\right)_{n_2}}{n_1! n_2! (c)_{n_1+n_2}}
z_1^{n_1} z_2^{n_2}.
\end{eqnarray}

The additional PDE reads:
\begin{eqnarray}
\theta _1 \theta _2\Phi_2(\vec{\gamma};\vec{\sigma};z_1,z_2)
=
(\frac{b_2 z_2}{z_1-z_2}
\theta _1
-\frac{b_1 z_1}{z_1-z_2}
\theta _2)
\Phi_2(\vec{\gamma};\vec{\sigma};z_1,z_2).
\end{eqnarray}

\begin{eqnarray}
A_{b_1,\Phi_2}
&=&
\frac{z_1}{b_1+b_2-c+1}+1,
\\
B_{b_1,\Phi_2}
&=&
-\frac{1}{b_1+b_2-c+1},
\\
C_{b_1,\Phi_2}
&=&
-\frac{z_1}{z_2 \left(b_1+b_2-c+1\right)},
\\
D_{b_1,\Phi_2}
&=&
0,
\end{eqnarray}

\begin{eqnarray}
A_{b_2,\Phi_2}
&=&
A_{b_1,\Phi_2}(b_1\leftrightarrow b_2,z_1\leftrightarrow z_2),
\\
B_{b_2,\Phi_2}
&=&
C_{b_1,\Phi_2}(b_1\leftrightarrow b_2,z_1\leftrightarrow z_2),
\\
C_{b_2,\Phi_2}
&=&
B_{b_1,\Phi_2}(b_1\leftrightarrow b_2,z_1\leftrightarrow z_2),
\\
D_{b_2,\Phi_2}
&=&
0,
\end{eqnarray}

\begin{eqnarray}
A_{c,\Phi_2}
&=&
\frac{1-c}{b_1+b_2-c+1},
\\
B_{c,\Phi_2}
&=&
-\frac{1-c}{z_1 \left(b_1+b_2-c+1\right)},
\\
C_{c,\Phi_2}
&=&
-\frac{1-c}{z_2 \left(b_1+b_2-c+1\right)},
\\
D_{c,\Phi_2}
&=&
0.
\end{eqnarray}

\boldmath
\subsubsection{Function $\left.\Phi_3(b,c,z_1,z_2\right)$}
\unboldmath

\begin{eqnarray}
\left.\Phi _3\text{(b,c,}z_1,z_2\right)
=\sum_{n_1,n_2}
\frac{(b)_{n_1}}{n_1! n_2! (c)_{n_1+n_2}}
z_1^{n_1} z_2^{n_2}.
\end{eqnarray}

The additional PDE reads:
\begin{eqnarray}
\theta _1 \theta _2\Phi_3(\vec{\gamma};\vec{\sigma};z_1,z_2)
=
(\frac{z_2}{z_1}
\theta _1
-b
\theta _2)
\Phi_3(\vec{\gamma};\vec{\sigma};z_1,z_2).
\end{eqnarray}

\begin{eqnarray}
A_{b,\Phi_3}
&=&
1,
\\
B_{b,\Phi_3}
&=&
0,
\\
C_{b,\Phi_3}
&=&
-\frac{z_1}{z_2},
\\
D_{b,\Phi_3}
&=&
0,
\end{eqnarray}

\begin{eqnarray}
A_{c,\Phi_3}
&=&
0,
\\
B_{c,\Phi_3}
&=&
0,
\\
C_{c,\Phi_3}
&=&
\frac{c-1}{z_2},
\\
D_{c,\Phi_3}
&=&
0.
\end{eqnarray}

\boldmath
\subsubsection{Function $\left.\Psi_1(a,b,c_1,c_2,z_1,z_2\right)$}
\unboldmath

\begin{eqnarray}
\left.\Psi _1\text{(a,b,}c_1,c_{2,}z_1,z_2\right)
=\sum_{n_1,n_2}
\frac{(a)_{n_1+n_2} (b)_{n_1}}{n_1! n_2! \left(c_1\right)_{n_1} \left(c_2\right)_{n_2}}
z_1^{n_1} z_2^{n_2},
\end{eqnarray}

\begin{eqnarray}
A_{a,\Psi_1}
&=&
\frac{b z_1}{a-c_1+1}+\frac{z_2}{a-c_2+1}+1,
\\
B_{a,\Psi_1}
&=&
\frac{\frac{z_2 \left(-2 a+c_1+c_2-2\right)}{\left(a-c_2+1\right) \left(a-c_1-c_2+2\right)}+z_1-1}{a-c_1+1},
\\
C_{a,\Psi_1}
&=&
\frac{\frac{b z_1 \left(-2 a+c_1+c_2-2\right)}{\left(a-c_1+1\right) \left(a-c_1-c_2+2\right)}-1}{a-c_2+1},
\\
D_{a,\Psi_1}
&=&
-\frac{\left(z_1-1\right) \left(2 a-c_1-c_2+2\right)}{\left(a-c_1+1\right) \left(a-c_2+1\right) \left(a-c_1-c_2+2\right)},
\end{eqnarray}

\begin{eqnarray}
A_{b,\Psi_1}
&=&
\frac{a z_1}{b-c_1+1}+1,
\\
B_{b,\Psi_1}
&=&
\frac{z_1-1}{b-c_1+1},
\\
C_{b,\Psi_1}
&=&
\frac{z_1}{b-c_1+1},
\\
D_{b,\Psi_1}
&=&
0,
\end{eqnarray}

\begin{eqnarray}
A_{c_1,\Psi_1}
&=&
\frac{\left(c_1-1\right) \left(a+b-c_1+1\right)}{\left(a-c_1+1\right) \left(-b+c_1-1\right)},
\\
B_{c_1,\Psi_1}
&=&
\frac{\left(c_1-1\right) \left(\left(z_1-1\right) \left(a-c_1-c_2+2\right)-z_2\right)}{z_1 \left(a-c_1+1\right) \left(a-c_1-c_2+2\right) \left(-b+c_1-1\right)},
\\
C_{c_1,\Psi_1}
&=&
\frac{b \left(c_1-1\right)}{\left(-a+c_1-1\right) \left(-a+c_1+c_2-2\right) \left(b-c_1+1\right)},
\\
D_{c_1,\Psi_1}
&=&
\frac{\left(c_1-1\right) \left(z_1-1\right)}{z_1 \left(a-c_1+1\right) \left(a-c_1-c_2+2\right) \left(b-c_1+1\right)},
\end{eqnarray}

\begin{eqnarray}
A_{c_2,\Psi_2}
&=&
A_{c_1,\Psi_2}(c_1\leftrightarrow c_2,z_1\leftrightarrow z_2),
\\
B_{c_2,\Psi_2}
&=&
C_{c_1,\Psi_2}(c_1\leftrightarrow c_2,z_1\leftrightarrow z_2),
\\
C_{c_2,\Psi_2}
&=&
B_{c_1,\Psi_2}(c_1\leftrightarrow c_2,z_1\leftrightarrow z_2),
\\
D_{c_2,\Psi_2}
&=&
D_{c_1,\Psi_2}(c_1\leftrightarrow c_2,z_1\leftrightarrow z_2).
\end{eqnarray}

\boldmath
\subsubsection{Function $\left.\Psi_2(a,c_1,c_2,z_1,z_2\right)$}
\unboldmath

\begin{eqnarray}
\left.\Psi _2\text{(a,}c_1,c_{2,}z_1,z_2\right)
=\sum_{n_1,n_2}
\frac{(a)_{n_1+n_2}}{n_1! n_2! \left(c_1\right)_{n_1} \left(c_2\right)_{n_2}}
z_1^{n_1} z_2^{n_2},
\end{eqnarray}

\begin{eqnarray}
A_{a,\Psi_2}
&=&
\frac{z_1}{a-c_1+1}+\frac{z_2}{a-c_2+1}+1,
\\
B_{a,\Psi_2}
&=&
\frac{\frac{z_2 \left(-2 a+c_1+c_2-2\right)}{\left(a-c_2+1\right) \left(a-c_1-c_2+2\right)}-1}{a-c_1+1},
\\
C_{a,\Psi_2}
&=&
\frac{\frac{z_1 \left(-2 a+c_1+c_2-2\right)}{\left(a-c_1+1\right) \left(a-c_1-c_2+2\right)}-1}{a-c_2+1},
\\
D_{a,\Psi_2}
&=&
-\frac{-2 (a+1)+c_1+c_2}{\left(a-c_1+1\right) \left(a-c_2+1\right) \left(a-c_1-c_2+2\right)},
\end{eqnarray}

\begin{eqnarray}
A_{c_1,\Psi_2}
&=&
\frac{a}{-a+c_1-1}+1,
\\
B_{c_1,\Psi_2}
&=&
\frac{\left(c_1-1\right) \left(a-c_1-c_2+z_2+2\right)}{z_1 \left(a-c_1+1\right) \left(a-c_1-c_2+2\right)},
\\
C_{c_1,\Psi_2}
&=&
\frac{c_1-1}{\left(-a+c_1-1\right) \left(-a+c_1+c_2-2\right)},
\\
D_{c_1,\Psi_2}
&=&
-\frac{c_1-1}{z_1 \left(-a+c_1-1\right) \left(-a+c_1+c_2-2\right)},
\end{eqnarray}

\begin{eqnarray}
A_{c_2,\Psi_2}
&=&
A_{c_1,\Psi_2}(c_1\leftrightarrow c_2,z_1\leftrightarrow z_2),
\\
B_{c_2,\Psi_2}
&=&
C_{c_1,\Psi_2}(c_1\leftrightarrow c_2,z_1\leftrightarrow z_2),
\\
C_{c_2,\Psi_2}
&=&
B_{c_1,\Psi_2}(c_1\leftrightarrow c_2,z_1\leftrightarrow z_2),
\\
D_{c_2,\Psi_2}
&=&
D_{c_1,\Psi_2}(c_1\leftrightarrow c_2,z_1\leftrightarrow z_2).
\end{eqnarray}

\boldmath
\subsubsection{Function $\left.\Theta_1(a_1,a_2,b,c,z_1,z_2\right)$}
\unboldmath

\begin{eqnarray}
\Theta _1\left(a_1,a_2\text{,b,c,}z_1,z_2\right)
=\sum_{n_1,n_2}
\frac{\left(a_1\right)_{n_1} \left(a_2\right)_{n_2} (b)_{n_1}}{n_1! n_2! (c)_{n_1+n_2}}
z_1^{n_1} z_2^{n_2},
\end{eqnarray}

\begin{eqnarray}
A_{a_1,\Theta_1}
&=&
\frac{b z_1}{a_1+a_2-c+1}+1,
\\
B_{a_1,\Theta_1}
&=&
-\frac{1-z_1}{a_1+a_2-c+1},
\\
C_{a_1,\Theta_1}
&=&
-\frac{b z_1}{z_2 \left(a_1+a_2-c+1\right)},
\\
D_{a_1,\Theta_1}
&=&
-\frac{z_1}{z_2 \left(a_1+a_2-c+1\right)},
\end{eqnarray}

\begin{eqnarray}
A_{a_2,\Theta_1}
&=&
-\frac{\left(a_2+b-c+1\right) \left(-a_2+c-z_2-1\right)-a_1 \left(a_2+b-c+z_2+1\right)}{\left(a_1+a_2-c+1\right) \left(a_2+b-c+1\right)},
\\
B_{a_2,\Theta_1}
&=&
\frac{\left(z_1-1\right) z_2}{z_1 \left(a_1+a_2-c+1\right) \left(a_2+b-c+1\right)},
\\
C_{a_2,\Theta_1}
&=&
-\frac{a_1+a_2+b-c+1}{\left(a_1+a_2-c+1\right) \left(a_2+b-c+1\right)},
\\
D_{a_2,\Theta_1}
&=&
-\frac{1}{\left(a_1+a_2-c+1\right) \left(a_2+b-c+1\right)},
\end{eqnarray}

\begin{eqnarray}
A_{b,\Theta_1}
&=&
\frac{a_1 z_1}{a_2+b-c+1}+1,
\\
B_{b,\Theta_1}
&=&
\frac{z_1-1}{a_2+b-c+1},
\\
C_{b,\Theta_1}
&=&
-\frac{a_1 z_1}{z_2 \left(a_2+b-c+1\right)},
\\
D_{b,\Theta_1}
&=&
-\frac{z_1}{z_2 \left(a_2+b-c+1\right)},
\end{eqnarray}

\begin{eqnarray}
A_{c,\Theta_1}
&=&
-\frac{(c-1) \left(a_1+a_2+b-c+1\right)}{\left(a_1+a_2-c+1\right) \left(a_2+b-c+1\right)},
\\
B_{c,\Theta_1}
&=&
-\frac{(c-1) \left(z_1-1\right)}{z_1 \left(a_1+a_2-c+1\right) \left(a_2+b-c+1\right)},
\\
C_{c,\Theta_1}
&=&
\frac{(c-1) \left(a_1+a_2+b-c+1\right)}{z_2 \left(a_1+a_2-c+1\right) \left(a_2+b-c+1\right)},
\\
D_{c,\Theta_1}
&=&
\frac{c-1}{z_2 \left(a_1+a_2-c+1\right) \left(a_2+b-c+1\right)}.
\end{eqnarray}

\boldmath
\subsubsection{Function $\left.\Theta_2(a,b,c,z_1,z_2\right)$}
\unboldmath

\begin{eqnarray}
\left.\Theta _2\text{(a,b,c,}z_1,z_2\right)
=\sum_{n_1,n_2}
\frac{(a)_{n_1} (b)_{n_1}}{n_1! n_2! (c)_{n_1+n_2}}
z_1^{n_1} z_2^{n_2},
\end{eqnarray}

\begin{eqnarray}
A_{a,\Theta_2}
&=&
1,
\\
B_{a,\Theta_2}
&=&
0,
\\
C_{a,\Theta_2}
&=&
-\frac{b z_1}{z_2},
\\
D_{a,\Theta_2}
&=&
-\frac{z_1}{z_2},
\end{eqnarray}

\begin{eqnarray}
A_{b,\Theta_2}
&=&
1,
\\
B_{b,\Theta_2}
&=&
0,
\\
C_{b,\Theta_2}
&=&
-\frac{a z_1}{z_2},
\\
D_{b,\Theta_2}
&=&
-\frac{z_1}{z_2},
\end{eqnarray}

\begin{eqnarray}
A_{c,\Theta_2}
&=&
0,
\\
B_{c,\Theta_2}
&=&
0,
\\
C_{c,\Theta_2}
&=&
\frac{c-1}{z_2},
\\
D_{c,\Theta_2}
&=&
0.
\end{eqnarray}

\boldmath
\subsubsection{Function $\left.\Gamma_1(a,b_1,b_2,z_1,z_2\right)$}
\unboldmath

\begin{eqnarray}
\left. \Gamma _1\text{(a,}b_1,b_2,z_1,z_2\right)
=\sum_{n_1,n_2}
\frac{(a)_{n_1} \left(b_1\right)_{n_2-n_1} \left(b_2\right)_{n_1-n_2}}{n_1! n_2!}
z_1^{n_1} z_2^{n_2}.
\end{eqnarray}

The additional PDE reads:
\begin{eqnarray}
\theta _1 \theta _2\Gamma_1(\vec{\gamma};z_1,z_2)
=
(a z_1 z_2
+
z_1 z_2
\theta _1)
\Gamma_1(\vec{\gamma};z_1,z_2).
\end{eqnarray}

\begin{eqnarray}
A_{a,\Gamma_1}
&=&
\frac{z_1 \left(z_2-b_2\right)}{a+b_1}+1,
\\
B_{a,\Gamma_1}
&=&
-\frac{z_1+1}{a+b_1},
\\
C_{a,\Gamma_1}
&=&
\frac{z_1}{a+b_1},
\\
D_{a,\Gamma_1}
&=&
0,
\end{eqnarray}

\begin{eqnarray}
A_{b_1,\Gamma_1}
&=&
\frac{b_1 \left(a+b_1+b_2-z_2\right)}{\left(b_1+b_2\right) \left(a+b_1\right)},
\\
B_{b_1,\Gamma_1}
&=&
\frac{b_1 \left(z_1+1\right)}{\left(b_1+b_2\right) z_1 \left(a+b_1\right)},
\\
C_{b_1,\Gamma_1}
&=&
-\frac{b_1}{\left(b_1+b_2\right) \left(a+b_1\right)},
\\
D_{b_1,\Gamma_1}
&=&
0,
\end{eqnarray}

\begin{eqnarray}
A_{b_2,\Gamma_1}
&=&
\frac{b_2}{b_1+b_2},
\\
B_{b_2,\Gamma_1}
&=&
0,
\\
C_{b_2,\Gamma_1}
&=&
\frac{b_2}{\left(b_1+b_2\right) z_2},
\\
D_{b_2,\Gamma_1}
&=&
0.
\end{eqnarray}

\boldmath
\subsubsection{Function $\left.\Gamma_2(b_1,b_2,z_1,z_2\right)$}
\unboldmath

\begin{eqnarray}
\Gamma _2\left(b_1,b_2,z_1,z_2\right)
=\sum_{n_1,n_2}
\frac{\left(b_1\right)_{n_2-n_1} \left(b_2\right)_{n_1-n_2}}{n_1! n_2!}
z_1^{n_1} z_2^{n_2}.
\end{eqnarray}

The additional PDE reads:
\begin{eqnarray}
\theta _1 \theta _2\Gamma_2(\vec{\gamma};z_1,z_2)
=
z_1 z_2\Gamma_2(\vec{\gamma};z_1,z_2).
\end{eqnarray}

\begin{eqnarray}
A_{b_1,\Gamma_2}
&=&
\frac{b_1}{b_1+b_2},
\\
B_{b_1,\Gamma_2}
&=&
\frac{b_1}{\left(b_1+b_2\right) z_1},
\\
C_{b_1,\Gamma_2}
&=&
0,
\\
D_{b_1,\Gamma_2}
&=&
0,
\end{eqnarray}

\begin{eqnarray}
A_{b_2,\Gamma_2}
&=&
A_{b_1,\Gamma_2}(b_1\leftrightarrow b_2,z_1\leftrightarrow z_2),
\\
B_{b_2,\Gamma_2}
&=&
0,
\\
C_{b_2,\Gamma_2}
&=&
B_{b_1,\Gamma_2}(b_1\leftrightarrow b_2,z_1\leftrightarrow z_2),
\\
D_{b_2,\Gamma_2}
&=&
0.
\end{eqnarray}

\boldmath
\subsubsection{Function $\left.H_1(a,b,c,z_1,z_2\right)$}
\unboldmath

\begin{eqnarray}
\left. H_1\text{(a,b,c,}z_1,z_2\right)
=\sum_{n_1,n_2}
\frac{(a)_{n_1-n_2} (b)_{n_1+n_2}}{n_1! n_2! (c)_{n_1}}
z_1^{n_1} z_2^{n_2},
\end{eqnarray}

\begin{eqnarray}
A_{a,H_1^c}
&=&
\frac{a \left(a+4 b z_1+b-2 c+2\right)}{(a+b) (a+b-2 c+2)},
\\
B_{a,H_1^c}
&=&
\frac{2 a \left(2 z_1-1\right)}{(a+b) (a+b-2 c+2)},
\\
C_{a,H_1^c}
&=&
\frac{a \left(\frac{4 z_1}{a+b-2 c+2}+\frac{1}{z_2}\right)}{a+b},
\\
D_{a,H_1^c}
&=&
-\frac{2 a}{z_2 (a+b) (a+b-2 c+2)},
\end{eqnarray}

\begin{eqnarray}
A_{b,H_1^c}
&=&
-\frac{2 (b+1) z_2 z_1 (a+3 b-2 c+2)}{(a+b) (b-c+1) (a+b-2 c+2)}
+\frac{a z_1}{b-c+1}+\frac{a+b-z_2}{a+b},
\\
B_{b,H_1^c}
&=&
\frac{-\frac{\left(2 z_1-1\right) z_2 (a+3 b-2 c+2)}{(a+b) (a+b-2 c+2)}+z_1-1}{b-c+1},
\\
C_{b,H_1^c}
&=&
-\frac{(a+b-2 c+2) \left(z_1 (a+b)+b-c+1\right)+2 z_1 z_2 (a+3 b-2 c+2)}{(a+b) (b-c+1) (a+b-2 c+2)},
\\
D_{b,H_1^c}
&=&
\frac{a+3 b-2 c+2}{(a+b) (b-c+1) (a+b-2 c+2)},
\end{eqnarray}

\begin{eqnarray}
A_{c,H_1^c}
&=&
-\frac{(c-1) \left(a^2+a (3 b-4 c+5)+2 (2 b-2 c+3) (b-c+1)-2 b z_2\right)}{(b-c+1) (a+b-2 c+2) (a+b-2 c+3)},
\\
B_{c,H_1^c}
&=&
\frac{(c-1) \left(-z_1 \left(a+3 b-4 c-2 z_2+5\right)+a+2 b-3 c-z_2+4\right)}{z_1 (b-c+1) (a+b-2 c+2) (a+b-2 c+3)},
\\
C_{c,H_1^c}
&=&
\frac{(c-1) \left(a-b+2 z_2+1\right)}{(b-c+1) (a+b-2 c+2) (a+b-2 c+3)},
\\
D_{c,H_1^c}
&=&
\frac{(c-1) \left(b-c-z_2+1\right)}{z_1 z_2 (b-c+1) (a+b-2 c+2) (a+b-2 c+3)}.
\end{eqnarray}

\boldmath
\subsubsection{Function $\left.H_2(a,b,c,d,z_1,z_2\right)$}
\unboldmath

\begin{eqnarray}
\left. H_2\text{(a,b,c,d,}z_1,z_2\right)
=\sum_{n_1,n_2}
\frac{(a)_{n_1-n_2} (b)_{n_1} (c)_{n_2}}{n_1! n_2! (d)_{n_1}}
z_1^{n_1} z_2^{n_2},
\end{eqnarray}

\begin{eqnarray}
A_{a,H_2^c}
&=&
\frac{a \left(a+b z_1+c-d+1\right)}{(a+c) (a+c-d+1)},
\\
B_{a,H_2^c}
&=&
\frac{a \left(z_1-1\right)}{(a+c) (a+c-d+1)},
\\
C_{a,H_2^c}
&=&
\frac{a}{z_2 (a+c)},
\\
D_{a,H_2^c}
&=&
-\frac{a}{z_2 (a+c) (a+c-d+1)},
\end{eqnarray}

\begin{eqnarray}
A_{b,H_2^c}
&=&
\frac{a z_1}{b-d+1}+1,
\\
B_{b,H_2^c}
&=&
\frac{z_1-1}{b-d+1},
\\
C_{b,H_2^c}
&=&
-\frac{z_1}{b-d+1},
\\
D_{b,H_2^c}
&=&
0,
\end{eqnarray}

\begin{eqnarray}
A_{c,H_2^c}
&=&
1-\frac{z_2 \left(a+b z_1+c-d+1\right)}{(a+c) (a+c-d+1)},
\\
B_{c,H_2^c}
&=&
-\frac{\left(z_1-1\right) z_2}{(a+c) (a+c-d+1)},
\\
C_{c,H_2^c}
&=&
-\frac{1}{a+c},
\\
D_{c,H_2^c}
&=&
\frac{1}{(a+c) (a+c-d+1)},
\end{eqnarray}

\begin{eqnarray}
A_{d,H_2^c}
&=&
-\frac{(d-1) (a+b+c-d+1)}{(b-d+1) (a+c-d+1)},
\\
B_{d,H_2^c}
&=&
-\frac{(d-1) \left(z_1-1\right)}{z_1 (b-d+1) (a+c-d+1)},
\\
C_{d,H_2^c}
&=&
0,
\\
D_{d,H_2^c}
&=&
\frac{d-1}{z_1 z_2 (b-d+1) (a+c-d+1)}.
\end{eqnarray}

\boldmath
\subsubsection{Function $\left.H_3(a,b,c,z_1,z_2\right)$}
\unboldmath

\begin{eqnarray}
\left. H_3\text{(a,b,c,}z_1,z_2\right)
=\sum_{n_1,n_2}
\frac{(a)_{n_1-n_2} (b)_{n_1}}{n_1! n_2! (c)_{n_1}}
z_1^{n_1} z_2^{n_2},
\end{eqnarray}

\begin{eqnarray}
A_{a,H_3^c}
&=&
0,
\\
B_{a,H_3^c}
&=&
0,
\\
C_{a,H_3^c}
&=&
\frac{a}{z_2},
\\
D_{a,H_3^c}
&=&
0,
\end{eqnarray}

\begin{eqnarray}
A_{b,H_3^c}
&=&
\frac{a z_1}{b-c+1}+1,
\\
B_{b,H_3^c}
&=&
\frac{z_1-1}{b-c+1},
\\
C_{b,H_3^c}
&=&
-\frac{z_1}{b-c+1},
\\
D_{b,H_3^c}
&=&
0,
\end{eqnarray}

\begin{eqnarray}
A_{c,H_3^c}
&=&
\frac{b}{-b+c-1}+1,
\\
B_{c,H_3^c}
&=&
0,
\\
C_{c,H_3^c}
&=&
0,
\\
D_{c,H_3^c}
&=&
\frac{c-1}{z_1 z_2 (b-c+1)}.
\end{eqnarray}

\boldmath
\subsubsection{Function $\left.H_4(a,b,c,z_1,z_2\right)$}
\unboldmath

\begin{eqnarray}
\left. H_4\text{(a,b,c,}z_1,z_2\right)
=\sum_{n_1,n_2}
\frac{(a)_{n_1-n_2} (b)_{n_2}}{n_1! n_2! (c)_{n_1}}
z_1^{n_1} z_2^{n_2},
\end{eqnarray}

\begin{eqnarray}
A_{a,H_4^c}
&=&
\frac{a \left(a+b-c+z_1+1\right)}{(a+b) (a+b-c+1)},
\\
B_{a,H_4^c}
&=&
-\frac{a}{(a+b) (a+b-c+1)},
\\
C_{a,H_4^c}
&=&
\frac{a}{z_2 (a+b)},
\\
D_{a,H_4^c}
&=&
-\frac{a}{z_2 (a+b) (a+b-c+1)},
\end{eqnarray}

\begin{eqnarray}
A_{b,H_4^c}
&=&
1-\frac{z_2 \left(a+b-c+z_1+1\right)}{(a+b) (a+b-c+1)},
\\
B_{b,H_4^c}
&=&
\frac{z_2}{(a+b) (a+b-c+1)},
\\
C_{b,H_4^c}
&=&
-\frac{1}{a+b},
\\
D_{b,H_4^c}
&=&
\frac{1}{(a+b) (a+b-c+1)},
\end{eqnarray}

\begin{eqnarray}
A_{c,H_4^c}
&=&
\frac{1-c}{a+b-c+1},
\\
B_{c,H_4^c}
&=&
\frac{c-1}{z_1 (a+b-c+1)},
\\
C_{c,H_4^c}
&=&
0,
\\
D_{c,H_4^c}
&=&
\frac{c-1}{z_1 z_2 (a+b-c+1)}.
\end{eqnarray}

\boldmath
\subsubsection{Function $\left.H_5(a,b,z_1,z_2\right)$}
\unboldmath

\begin{eqnarray}
\left. H_5\text{(a,b,}z_1,z_2\right)
=\sum_{n_1,n_2}
\frac{(a)_{n_1-n_2}}{n_1! n_2! (b)_{n_1}}
z_1^{n_1} z_2^{n_2},
\end{eqnarray}

\begin{eqnarray}
A_{a,H_5^c}
&=&
0,
\\
B_{a,H_5^c}
&=&
0,
\\
C_{a,H_5^c}
&=&
\frac{a}{z_2},
\\
D_{a,H_5^c}
&=&
0,
\end{eqnarray}

\begin{eqnarray}
A_{b,H_5^c}
&=&
0,
\\
B_{b,H_5^c}
&=&
0,
\\
C_{b,H_5^c}
&=&
0,
\\
D_{b,H_5^c}
&=&
\frac{b-1}{z_1 z_2}.
\end{eqnarray}

\boldmath
\subsubsection{Function $\left.H_6(a,b,z_1,z_2\right)$}
\unboldmath

\begin{eqnarray}
H_6\left(a,b,z_1,z_2\right)
=\sum_{n_1,n_2}
\frac{(a)_{2 n_1+n_2}}{n_1! n_2! (b)_{n_1+n_2}}
z_1^{n_1} z_2^{n_2}.
\end{eqnarray}

The additional PDE reads:
\begin{eqnarray}
(-z_2
\theta _1
+
a z_1
\theta _2
+
z_1
\theta _2{}^2
+
2 z_1
\theta _1\theta _2)
H_6(\vec{\gamma};\vec{\sigma};z_1,z_2)
=0.
\end{eqnarray}

\begin{eqnarray}
A_{a,H_6^c}
&=&
\frac{2 (a+1) z_1+a-b+z_2+1}{a-b+1},
\\
B_{a,H_6^c}
&=&
\frac{4 z_1-1}{a-b+1},
\\
C_{a,H_6^c}
&=&
-\frac{z_1 \left(a-2 z_2\right)+z_2}{z_2 (a-b+1)},
\\
D_{a,H_6^c}
&=&
0,
\end{eqnarray}

\begin{eqnarray}
A_{b,H_6^c}
&=&
\frac{(b-1) \left(2 a z_1+z_2\right)}{z_2 (-a+b-1)},
\\
B_{b,H_6^c}
&=&
-\frac{(b-1) \left(4 z_1-1\right)}{z_2 (a-b+1)},
\\
C_{b,H_6^c}
&=&
\frac{(b-1) \left(z_2-z_1 \left(a-2 b+2 z_2+3\right)\right)}{z_2^2 (a-b+1)},
\\
D_{b,H_6^c}
&=&
0.
\end{eqnarray}

\boldmath
\subsubsection{Function $\left.H_7(a,c,d,z_1,z_2\right)$}
\unboldmath

\begin{eqnarray}
\left. H_7\text{(a,c,d,}z_1,z_2\right)
=\sum_{n_1,n_2}
\frac{(a)_{2 n_1+n_2}}{n_1! n_2! (c)_{n_1} (d)_{n_2}}
z_1^{n_1} z_2^{n_2},
\end{eqnarray}

\begin{eqnarray}
A_{a,H_7^c}
&=&
\frac{4 (a+1) z_2 z_1 (-2 a+2 c+d-3)}{(a-2 c+2) (a-d+1) (a-2 c-d+3)}
+\frac{4 (a+1) z_1}{a-2 c+2}
\nn\\
&&
+\frac{a-d+z_2+1}{a-d+1},
\\
B_{a,H_7^c}
&=&
\frac{2 \left(-\frac{\left(4 z_1+1\right) z_2 (2 a-2 c-d+3)}{(a-d+1) (a-2 c-d+3)}+4 z_1-1\right)}{a-2 c+2},
\\
C_{a,H_7^c}
&=&
\frac{4 z_1 \left(z_2 (-2 a+2 c+d-3)+a (-a+d-3)+2 c+d-3\right)}{(a-2 c+2) (a-d+1) (a-2 c-d+3)}
\nn\\
&&+\frac{1}{-a+d-1},
\\
D_{a,H_7^c}
&=&
-\frac{2 \left(4 z_1-1\right) (2 a-2 c-d+3)}{(a-2 c+2) (a-d+1) (a-2 c-d+3)},
\end{eqnarray}

\begin{eqnarray}
A_{c,H_7^c}
&=&
\frac{2 a (c-1) z_2 \left(a-2 c+z_2+2\right)}{(a-2 c+2) (a-2 c+3) (a-2 c-d+3) (a-2 c-d+4)}
\nn\\
&&
-\frac{2 (c-1) (2 a-2 c+3)}{(a-2 c+2) (a-2 c+3)},
\\
B_{c,H_7^c}
&=&
\frac{4 (c-1) z_2}{(a-2 c+3) (a-2 c-d+3) (a-2 c-d+4)}
\nn\\
&&
+\frac{(c-1) z_2 \left(3 a-2 (3 c+d-5)+\left(4 z_1+1\right) z_2\right)}{z_1 (a-2 c+2) (a-2 c+3) (a-2 c-d+3) (a-2 c-d+4)}
\nn\\
&&
-\frac{(c-1) \left(4 z_1-1\right)}{z_1 (a-2 c+2) (a-2 c+3)},
\\
C_{c,H_7^c}
&=&
\frac{4 (a+1) (c-1)}{(a-2 c+2) (a-2 c-d+3) (a-2 c-d+4)}
\nn\\
&&
-\frac{2 (c-1) \left(d-z_2\right) \left(2 a-2 c+z_2+3\right)}{(a-2 c+2) (a-2 c+3) (a-2 c-d+3) (a-2 c-d+4)},
\\
D_{c,H_7^c}
&=&
\frac{(c-1) \left(4 z_1-1\right) \left(2 a-4 c-d+z_2+6\right)}{z_1 (a-2 c+2) (a-2 c+3) (a-2 c-d+3) (a-2 c-d+4)},
\end{eqnarray}

\begin{eqnarray}
A_{d,H_7^c}
&=&
\frac{(d-1) \left(4 a z_1-a+2 c+d-3\right)}{(a-d+1) (a-2 c-d+3)},
\\
B_{d,H_7^c}
&=&
\frac{2 (d-1) \left(4 z_1+1\right)}{(a-d+1) (a-2 c-d+3)},
\\
C_{d,H_7^c}
&=&
\frac{(d-1) \left(4 z_1 \left(a-d+z_2+2\right)+a-2 c-d+3\right)}{z_2 (a-d+1) (a-2 c-d+3)},
\\
D_{d,H_7^c}
&=&
\frac{2 (d-1) \left(4 z_1-1\right)}{z_2 (a-d+1) (a-2 c-d+3)}.
\end{eqnarray}

\boldmath
\subsubsection{Function $\left.H_8(a,b,z_1,z_2\right)$}
\unboldmath

\begin{eqnarray}
H_8\left(a,b,z_1,z_2\right)
=\sum_{n_1,n_2}
\frac{(a)_{2 n_1-n_2} (b)_{n_2-n_1}}{n_1! n_2!}
z_1^{n_1} z_2^{n_2}.
\end{eqnarray}

The additional PDE reads:
\begin{eqnarray}
(-a z_1 z_2
-2 z_1 z_2
\theta _1
+
z_1 z_2
\theta _2
+
\theta _1\theta _2)
H_8(\vec{\gamma};z_1,z_2)
=0.
\end{eqnarray}

\begin{eqnarray}
A_{a,H_8^c}
&=&
\frac{a \left(z_1 \left(z_2-2 (a+1)\right)+a+2 b\right)}{(a+b) (a+2 b)},
\\
B_{a,H_8^c}
&=&
-\frac{a \left(4 z_1+1\right)}{(a+b) (a+2 b)},
\\
C_{a,H_8^c}
&=&
\frac{a \left(a+2 b+2 z_1 z_2\right)}{z_2 (a+b) (a+2 b)},
\\
D_{a,H_8^c}
&=&
0,
\end{eqnarray}

\begin{eqnarray}
A_{b,H_8^c}
&=&
\frac{b \left(2 (a+b) (2 a+2 b+1)-z_2 \left(z_1 \left(z_2-2 a\right)+2 a+3 b+1\right)\right)}{(a+b) (a+2 b) (a+2 b+1)},
\\
B_{b,H_8^c}
&=&
\frac{b \left(4 z_1+1\right) \left(a+b+z_1 z_2\right)}{z_1 (a+b) (a+2 b) (a+2 b+1)},
\\
C_{b,H_8^c}
&=&
-\frac{b \left(3 a+4 b+2 z_1 z_2+1\right)}{(a+b) (a+2 b) (a+2 b+1)},
\\
D_{b,H_8^c}
&=&
0.
\end{eqnarray}

\boldmath
\subsubsection{Function $\left.H_9(a,b,c,z_1,z_2\right)$}
\unboldmath

\begin{eqnarray}
\left. H_9\text{(a,b,c,}z_1,z_2\right)
=\sum_{n_1,n_2}
\frac{(a)_{2 n_1-n_2} (b)_{n_2}}{n_1! n_2! (c)_{n_1}}
z_1^{n_1} z_2^{n_2},
\end{eqnarray}

\begin{eqnarray}
A_{a,H_9^c}
&=&
\frac{a \left(4 (a+1) z_1+a+b-2 c+2\right)}{(a+b) (a+b-2 c+2)},
\\
B_{a,H_9^c}
&=&
\frac{2 a \left(4 z_1-1\right)}{(a+b) (a+b-2 c+2)},
\\
C_{a,H_9^c}
&=&
\frac{a \left(\frac{1}{z_2}-\frac{4 z_1}{a+b-2 c+2}\right)}{a+b},
\\
D_{a,H_9^c}
&=&
-\frac{2 a}{z_2 (a+b) (a+b-2 c+2)},
\end{eqnarray}

\begin{eqnarray}
A_{b,H_9^c}
&=&
1-\frac{z_2 \left(4 a z_1+a+b-2 c+2\right)}{(a+b) (a+b-2 c+2)},
\\
B_{b,H_9^c}
&=&
\frac{2 \left(1-4 z_1\right) z_2}{(a+b) (a+b-2 c+2)},
\\
C_{b,H_9^c}
&=&
-\frac{a+b-2 c-4 z_1 z_2+2}{(a+b) (a+b-2 c+2)},
\\
D_{b,H_9^c}
&=&
\frac{2}{(a+b) (a+b-2 c+2)},
\end{eqnarray}

\begin{eqnarray}
A_{c,H_9^c}
&=&
\frac{2 (c-1) \left(a b-z_2 (2 a+b-2 c+3)\right)}{z_2 (a+b-2 c+2) (a+b-2 c+3)},
\\
B_{c,H_9^c}
&=&
\frac{(c-1) \left(4 z_1-1\right) \left(b-z_2\right)}{z_1 z_2 (a+b-2 c+2) (a+b-2 c+3)},
\\
C_{c,H_9^c}
&=&
-\frac{2 (c-1) \left(b-z_2\right)}{z_2 (a+b-2 c+2) (a+b-2 c+3)},
\\
D_{c,H_9^c}
&=&
\frac{(c-1) \left(a-2 c+z_2+3\right)}{z_1 z_2^2 (a+b-2 c+2) (a+b-2 c+3)}.
\end{eqnarray}

\boldmath
\subsubsection{Function $\left.H_{10}(a,c,z_1,z_2\right)$}
\unboldmath

\begin{eqnarray}
\left. H_{10}\text{(a,c,}z_1,z_2\right)
=\sum_{n_1,n_2}
\frac{(a)_{2 n_1-n_2}}{n_1! n_2! (c)_{n_1}}
z_1^{n_1} z_2^{n_2},
\end{eqnarray}

\begin{eqnarray}
A_{a,H_{10}^c}
&=&
0,
\\
B_{a,H_{10}^c}
&=&
0,
\\
C_{a,H_{10}^c}
&=&
\frac{a}{z_2},
\\
D_{a,H_{10}^c}
&=&
0,
\end{eqnarray}

\begin{eqnarray}
A_{c,H_{10}^c}
&=&
\frac{2 a (c-1)}{z_2},
\\
B_{c,H_{10}^c}
&=&
\frac{(c-1) \left(4 z_1-1\right)}{z_1 z_2},
\\
C_{c,H_{10}^c}
&=&
\frac{2-2 c}{z_2},
\\
D_{c,H_{10}^c}
&=&
\frac{(c-1) (a-2 c+3)}{z_1 z_2^2}.
\end{eqnarray}

\boldmath
\subsubsection{Function $\left.H_{11}(a,b,c,d,z_1,z_2\right)$}
\unboldmath

\begin{eqnarray}
\left. H_{11}\text{(a,b,c,d,}z_1,z_2\right)
=\sum_{n_1,n_2}
\frac{(a)_{n_1-n_2} (b)_{n_2} (c)_{n_2}}{n_1! n_2! (d)_{n_1}}
z_1^{n_1} z_2^{n_2},
\end{eqnarray}

\begin{eqnarray}
A_{a,H_{11}^c}
&=&
\frac{z_1 \left(a^2+a (b+2 c-d+1)+c (-b+c-d+1)\right)}{(a+c) (a+b-d+1) (a+c-d+1)}
\nn\\
&&
+\frac{b c z_1 (b-c+d-1)}{(a+b) (a+c) (a+b-d+1) (a+c-d+1)}
\nn\\
&&
+\frac{a (a+b+c)}{(a+b) (a+c)},
\\
B_{a,H_{11}^c}
&=&
\frac{a \left(\frac{c}{(a+b) (a+b-d+1)}-\frac{b}{(a+c) (a+c-d+1)}\right)}{b-c},
\\
C_{a,H_{11}^c}
&=&
\frac{a z_1 (2 a+b+c-d+1)}{(a+b) (a+c) (a+b-d+1) (a+c-d+1)}
\nn\\
&&
+\frac{a \left(z_2+1\right)}{z_2 (a+b) (a+c)},
\\
D_{a,H_{11}^c}
&=&
-\frac{a \left(z_2+1\right) (2 a+b+c-d+1)}{z_2 (a+b) (a+c) (a+b-d+1) (a+c-d+1)},
\end{eqnarray}

\begin{eqnarray}
A_{b,H_{11}^c}
&=&
1-\frac{c z_2 \left(a+b-d+z_1+1\right)}{(a+b) (a+b-d+1)},
\\
B_{b,H_{11}^c}
&=&
\frac{c z_2}{(a+b) (a+b-d+1)},
\\
C_{b,H_{11}^c}
&=&
-\frac{z_2 \left(a+b-d+z_1+1\right)+a+b-d+1}{(a+b) (a+b-d+1)},
\\
D_{b,H_{11}^c}
&=&
\frac{z_2+1}{(a+b) (a+b-d+1)},
\end{eqnarray}

\begin{eqnarray}
A_{c,H_{11}^c}
&=&
A_{b,H_{11}^c}(b\leftrightarrow c),
\\
B_{c,H_{11}^c}
&=&
B_{b,H_{11}^c}(b\leftrightarrow c),
\\
C_{c,H_{11}^c}
&=&
C_{b,H_{11}^c}(b\leftrightarrow c),
\\
D_{c,H_{11}^c}
&=&
D_{b,H_{11}^c}(b\leftrightarrow c),
\end{eqnarray}

\begin{eqnarray}
A_{d,H_{11}^c}
&=&
-\frac{(d-1) (a+b+c-d+1)}{(a+b-d+1) (a+c-d+1)},
\\
B_{d,H_{11}^c}
&=&
\frac{(d-1) (a+b+c-d+1)}{z_1 (a+b-d+1) (a+c-d+1)},
\\
C_{d,H_{11}^c}
&=&
\frac{1-d}{(a+b-d+1) (a+c-d+1)},
\\
D_{d,H_{11}^c}
&=&
\frac{(d-1) \left(z_2+1\right)}{z_1 z_2 (a+b-d+1) (a+c-d+1)}.
\end{eqnarray}

\end{document}